\documentclass[acs,twocolumn,amsmath,amssymb,reprint]{revtex4}
\usepackage{graphicx}
\usepackage{amsmath}
\usepackage{latexsym}
\usepackage{amsmath}
\usepackage{amssymb}
\usepackage{float}
\usepackage{color}
\usepackage{epstopdf}
\usepackage{hyperref}
\hypersetup{urlcolor=red,citecolor=blue}
\usepackage{bm}
\hypersetup{urlcolor=red,citecolor=blue}

\usepackage{epsfig}

\definecolor{g1}{rgb}{0.0,0.0,0.0}

\newcommand{\nds}[1]{C$_{3}$H$_{8}$}
\newcommand{\ada}[1]{C$_{10}$H$_{16}$}
\newcommand{\penta}[1]{C$_{26}$H$_{32}$}
\newcommand{\ndl}[1]{C$_{84}$H$_{64}$}
\newcommand{\pz}[1]{$p_z$}
\newcommand{\sptwo}[1]{$sp^2$}
\newcommand{\spthree}[1]{$sp^3$}
\newcommand{\csixty}[1]{$C_{60}$}

\begin{document}

\title{Hybrid atomic orbital basis from first principles: \\
Bottom-up mapping of self-energy correction to large covalent systems}
\author{Manoar Hossain, Joydev De, Joydeep Bhattacharjee}
\affiliation{National Institute of Science Education and Research, HBNI, Jatni,
Khurda, Bhubaneswar, 752050, Odisha, India}
\begin{abstract}

Construction of hybrid atomic orbitals is proposed as the approximate common eigen states of finite first moment matrices.  
Their hybridization and orientation can be a-priori tunned as per their anticipated neighbourhood.
Their  Wannier function counterparts constructed from the Kohn-Sham(KS) single particle states
constitute an orthonormal multi-orbital tight-binding(TB) basis resembling hybrid atomic-orbitals 
locked to their immediate atomic neighborhood, while spanning the subs-space of KS states.
The proposed basis thus not only renders predominantly single TB parameters  from first-principles
for each nearest neighbour bonds involving no more than two orbitals irrespective of their orientation,
but also facilitate an easy route for transfer of such TB parameters  across isostructural systems 
exclusively through mapping of neighbourhoods and projection of orbital charge centres.
With hybridized 2$s$,2$p$ and 3$s$,3$p$ valence electrons,
the spatial extent of self-energy correction(SEC) to TB parameters in the proposed basis  
are found to be  localized
mostly within the third nearest neighbourhood, thus allowing effective transfer of 
self-energy corrected TB parameters  from smaller reference systems 
to much larger target systems,
with nominal additional computational cost beyond that required for explicit computation of SEC in the reference systems.
The proposed approach promises inexpensive estimation of quasi-particle structure of large covalent systems with workable accuracy.
\end{abstract}
\maketitle
\subsection{Introduction}
\label{intro}
Setting a minimal TB basis for a given systems of atoms 
calls for appropriate orientation of orbitals at each atomic site in accordance with their immediate atomic neighbourhood,
so that the nearest neighbour interactions can be represented by the least number of orbitals.
In this direction, hybrid atomic orbitals have been used by quantum chemists
since their introduction\cite{pauling1931nature,slater1931directed} almost a century ago. 
Rational approaches for their 
construction\cite{doggett1965excited,del1963hybridization,mcweeny1960some,del1966bent,reed1988intermolecular}
over the last several decades have been primarily focussed on partitioning systems into substructures 
which are spanned by groups of hybrid orbitals, leading to unambiguous partitioning of electrons into
bonding orbitals and lone-pairs, and further into atomic orbitals. 
For such partitioning, notionally similar several approaches
\cite{del1963hybridization,mcweeny1968criteria,del1966bent,murrell1960construction,mayer1996atomic,mayer1995non}
have been proposed grossly based on the maximum overlap condition which in effect leads to localization
of orbitals within the chosen subspace of molecular orbitals. 
In these approaches, either the overlap matrix\cite{del1963hybridization,del1966bent}  
or the first-order density matrices\cite{mayer1996atomic,foster1980natural},
both of which are calculated typically in the basis of either the Slater type orbitals(STO)\cite{slater1930atomic} 
or the Gaussian type orbitals(GTO)\cite{hehre1969self,dunning1989gaussian},
are generally transformed into block diagonal forms each spanned by orbitals centered on nearest neighbour atoms.
The resultant hybrid orbitals involving atomic orbitals centred on more than one atoms\cite{reed1988intermolecular,zhan1993maximum} 
render unambiguous bonding orbitals and bond-orders,
while the ones like the  {\it natural hybrid orbitals}(NHO)\cite{foster1980natural} 
or the {\it effective atomic orbital}(EAO)\cite{mayer1995non}, which involve atomic orbitals of a single atom, 
describe the state of the orbitals of the atoms as they participate in bonds.
Hybird orbitals in the line of NHOs have been popularly constructed ab-initio at the HF level\cite{rives1980natural,gopinathan1988determination}.


A more explicit approach\cite{baxter1996molecular,kirkwood1977generalized} has been to construct the {\it generalized hybrid orbitals}(GHO) 
as combinations of STO with common Slater exponent and fixed position of nodes along bonds to assign their orientation.
Expedient to clarify that in this paper we refer to bonds simply as the linear connectivity between atoms 
which are primarily nearest neighbours if not mentioned specifically.
%
Much of these efforts were undertaken in aid to molecular mechanics calculation\cite{gao1998generalized,pu2004generalized} 
where the description of interactions between sub-structures eases with use of orbitals which are directed along bonds. 
Effective analytical models for such interactions have also been developed\cite{popov2020deductive} recently for 
inexpensive deductive computation of properties of bulk as well as clusters of $sp^x$ hybridized covalent systems.
%
Notably unlike the GHOs, the NHOs or the EAOs by construction may not be oriented exactly along the bonds.
In general for all such hybrid orbitals, their directed nature, maximal localization and orthonormality are not
guaranteed simultaneously by construction.
In a part of this work we explore simultaneity of these conditions in construction of hybrid atomic orbitals 
from first-principles proposed in this work.



Instead of overlap or density matrices, in this work we take recourse to the first moment matrices (FMM)
due to their direct correspondence to localization.
FMMs are known not to commute among  each other in more than one dimension if projected on to a finite subspace of orthonormal states.
We propose construction of hybrid atomic orbitals(HAO) as approximate eigen states of the FMMs 
within a finite subspace of Kohn-Sham (KS) states of isolated atoms. 
Orientation and  hybridization of the proposed orbitals can be a-priori naturalized as per their anticipated neighbourhood.
%
%
This substantially eases the effort of orientating them appropriately while
transferring them from isolated atoms to the real systems, which eventually eases the interpretation of 
elements of the Hamiltonian. 
An orthonormal set of localized Wannier orbitals resembling the HAOs 
is further constructed in the basis of  KS single particle states of the given system.
These Wannier orbitals, 
which we refer in this paper as the  {\it hybrid atomic Wannier orbitals} (HAWO),
constitute a multi-orbital tight-binding (TB) basis locked to their immediate atomic neighbourhood by construction,  
and render hopping parameters involving effectively only two orbitals per bond.
HAWOs thus offer easy transfer of the corresponding TB parameters to other iso-structural systems 
exclusively through mapping of neighbourhoods and projection of charge centres learned from HAOs.
Effective transfer of TB parameters is demonstrated in 
nano-ribbons of graphene and hexagonal boron-nitride, \csixty \\, and nano-diamonds and their silicon based counterparts.
In particular, we show in the HAWO basis that it is possible to effectively transfer 
self-energy(SE) correction(SEC) of single particle levels 
from smaller reference systems to much larger iso-structural systems through TB parameters
with minimal additional computational expense through the proposed mapping of multi-orbital TB parameters beyond the nearest neighbourhood.
%
\section{Methodological details}
\subsection{Construction of hybrid orbitals}
\label{mhao}
In a given direction, 
for example along $\hat{x}$, the most localized orbtials $\left\{ \phi \right\}$ 
 would diagonalize the corresponding FMM: 
\begin{equation}
X_{ij}=\langle \phi_i \mid x \mid \phi_j \rangle. 
\label{firstmom}
\end{equation}
%
This becomes clear by noting that the total spread
of a finite set of $N$ number of orbitals  along $\hat{x}$ 
is given by:
\begin{equation}
\Omega_x = \sum_ {i=1,N} \left[ \langle \phi_i | x^2 | \phi_i \rangle 
- |\langle \phi_i | x | \phi_i \rangle|^2 \right],
\end{equation}
which can be expressed as:
\begin{eqnarray}
\Omega_x&=& \sum_ {i=1,N}  \left( \sum^{\infty}_{j=1}  X_{ij}X_{ji} - X_{ii}X_{ii} \right) \nonumber \\
&=& \sum_ {i=1,N} \sum^{\infty}_{j\ne i} |X_{ij}|^2   \nonumber \\
&=& \sum_ {i=1,N}  \left( \sum^N_{j\ne i} |X_{ij}|^2 +   \sum^{\infty}_{j=N+1} |X_{ij}|^2 \right).
\label{omegaWF}
\end{eqnarray}
Diagonalization of $X$ in the $N\times N$ subspace would therefore sets the first term in Eqn.(\ref{omegaWF}) to zero,
leading to minimization of total spread. 
Notably, $X$ can be calculated directly as in Eqn.(\ref{firstmom})
only  for isolated systems well separated from their periodic images.
For periodic system with non-zero crystal momentum, computation of  $X$ 
would essentially involve evaluation of geometric phases\cite{berry1984quantal} of Bloch electrons 
evolved across the Brillouin Zone\cite{resta1994macroscopic,king1993theory}. 
Nevertheless, there 
exists therefore a unique set of orbitals  
which completely diagonalize $X$, 
and would also thereby have maximum localization along $\hat{x}$.
Similar unique sets exist for $\hat{y}$ and $\hat{z}$ directions as well.
%
However, the matrices $X$, $Y$ and $Z$, when projected into a finite subspace of orthonormal states, 
do not commute with each other in general unless mandated by symmetries.
This implies that a unique set of orbitals with maximum localization simultaneously
in all three orthogonal directions would not exist in general.
The same is true for Wannier functions (WF) in case of periodic systems with non-zero wave-vectors.
Numerically localized Wannier functions \cite{marzari1997maximally,bhattacharjee2006localized}
therefore are not be unique and the choice
of gauge used for their construction depends on the chosen criteria of localization. 


We chose to look for the  possibility to construct a set of localized orbitals which will be a reasonable 
compromise between the three unique sets of orbitals having maximum localization along the three orthogonal
directions.
%
We thus  resorted to the condition of simultaneous approximate joint diagonalization \cite{cardoso1996jacobi} 
of the three FMMs: $X$, $Y$ and $Z$. 
%
To compute such an approximate eigen sub-space of the three FMMs, we adopted an iterative scheme 
based on generalization of the Jacobi method of matrix diagonalization\cite{goldstine1959jacobi}, 
wherein,  off-diagonal elements are iteratively minimized by applying rotation of coordinates by
an optimally chosen angle. 
The extension of the method to more than one square matrices
irrespective of whether they are commuting or not, is based on a proposed\cite{cardoso1996jacobi} choice of angle 
of rotation leading to complex rotation matrix $U$ which has been proven\cite{cardoso1996jacobi}
to minimize the composite objective function defined as :
\begin{equation}
\mbox{off}(UXU^\dagger)+\mbox{off}(UYU^\dagger)+\mbox{off}(UZU^\dagger)
\label{obj}
\end{equation}
where $\mbox{off}(A)=\sum_{1 \le i \ne j \ge N} |A_{i j}|^2$ for an $N \times N$ matrix $A$. 
$N$ being the number of orthonormal states used to compute $X$, $Y$ and $Z$. 
$U$ is a product of all the $N(N-1)/2$ complex plane rotations, one each for each pairs of $\left(ij\right)$ for $i\ne j$.
For a given $\left(ij\right)$ the plane rotation $R(i,j)$ is an $N \times N$ identity matrix except for:
\begin{equation}
\left( \begin{array}{cc}
r_{ii} & r_{ij}  \\
r_{ji} & r_{jj}  \end{array} \right)
= \left( \begin{array}{cc}
c & \overline{s}  \\
-s & \overline{c}  \end{array} \right)
\label{rij}
\end{equation}
where $c,s \in C, |c|^2+|s|^2=1$.


It has been shown\cite{cardoso1996jacobi} that the objective function defined in Eqn.(\ref{obj})
is minimized if $U$ is a product of $R(i,j)$ matrices as shown in Eqn.(\ref{rij}) 
whose elements  are given as: 
\begin{equation}
c=\sqrt{\frac{x+r}{2r}}; 
s=\frac{y-iz}{\sqrt{2r(x+r)}}
\end{equation}
where 
\[ r=\sqrt{x^2+y^2+z^2} \]
and $[x,y,z]^\dagger$ being the eigen-vector corresponding to the highest eigen-value
of a 3$\times$3 matrix:
\begin{eqnarray}
G(i,j) =& &\mbox{Real}\left(h^\dagger(X,i,j)h(X,i,j)  \right) \nonumber \\
        &+&\mbox{Real}\left(h^\dagger(Y,i,j)h(Y,i,j)  \right)  \nonumber \\
        &+&\mbox{Real}\left(h^\dagger(Z,i,j)h(Z,i,j)  \right)  \nonumber
\end{eqnarray}
with:
\begin{equation}
h(A,i,j)=[a_{ii}-a_{jj},a_{ij}+a_{ji}, i(a_{ji}-a_{ij})].
\end{equation}


Notably, given the form of $R(i,j)$, for a rotated matrix $A'=R(i,j)AR^\dagger(i,j)$ 
corresponding to plane rotation for the $\left(ij\right)$-th pair of elements of $A$,
it is easily seen that  
$a'_{kk}=a_{kk}$ for $k \ne i$ and $k \ne  j$, leading to the invariance:
\[
\mbox{off}(A')+ |a'_{ii}|^2 + |a'_{jj}|^2= \mbox{off}(A) + |a_{ii}|^2 + |a_{jj}|^2.
\]
owing to preservation of norm in similarity transformation.
Therefore, minimizing $\mbox{off}(A')$ would naturally imply maximising $ |a'_{ii}|^2 + |a'_{jj}|^2$,
which further implies maximising $|a'_{ii} - a'_{jj}|^2$ since:
\[
2\left(|a'_{ii}|^2+|a'_{jj}|^2\right)=|a'_{ii}+a'_{jj}|^2+|a'_{ii}-a'_{jj}|^2
\]
and 
\[
a'_{ii}+a'_{jj}=a_{ii}+a_{jj}
\]
owing to invariance of trace under similarity transformation.
Therefore in our case the minimization of the objective function[Eqn.(\ref{obj})]
implies maximizing the separation between the charge centres of the $i$-th and the $j$-th orbitals,
which is thus similar to the principle of the Foster and Boys\cite{foster1960canonical} 
scheme of orbital localization. 
This becomes clear by rewriting  the total spread [Eqn.(\ref{omegaWF})] for $N$ orbital $\left\{ \phi_i,i=1,N \right\}$ as:
\begin{equation}
\Omega 
= \sum_{k=1,3} \sum_ {i=1,N}  \left( \sum^N_{j\ne i} |a^k_{ij}|^2 +   \sum^{\infty}_{j=N+1} |a^k_{ij}|^2 \right)
\label{omega}
\end{equation}
where $A^{k=1,2,3}=X,Y,Z$.
Eqn.(\ref{omega}) clearly suggests that minimization of the objective function in Eqn.(\ref{obj}) 
would minimizes the first term in  Eqn.(\ref{omega}), leading to minimization of the total spread. 
Eqn.\ref{omega} also suggests that the total spread will reduce with increasing number of states ($N$) in
the basis of which the first moment matrices are constructed.
%
%
%

\begin{figure}[t]
\includegraphics[scale=0.33]{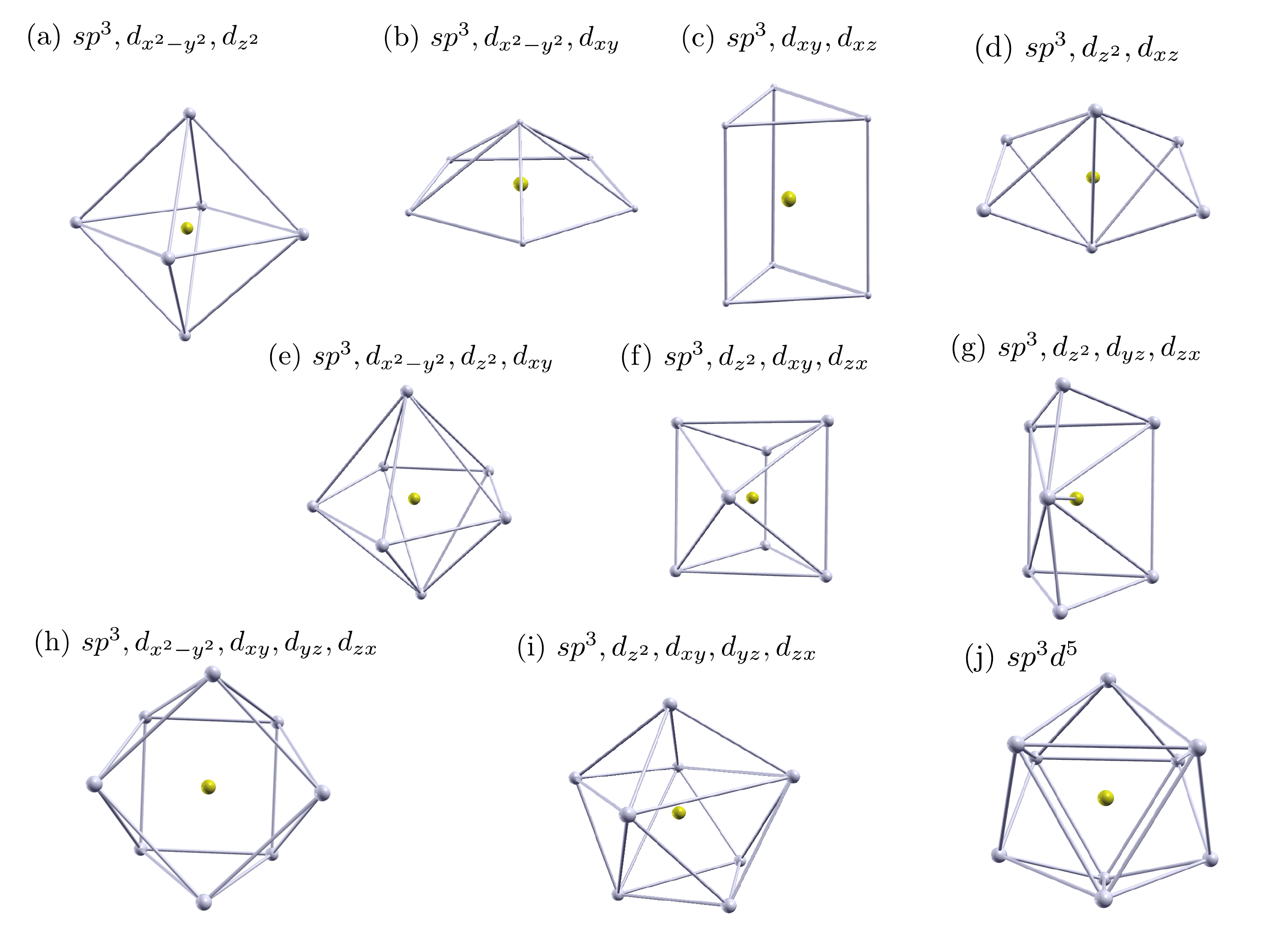}
\caption{Plots of charge centres (shown in gray) of the hybrid orbitals formed by the group of 
GTOs representing $3s, 3p$ and $3d$ orbitals of Ti (shown in yellow) 
constructed as per Ref.\cite{tatewaki1979systematic} . }
\label{polyhedra}
\end{figure}

We test the proposed approach first with FMMs computed in the basis of GTOs constructed for Ti with parameters 
from Ref.\cite{tatewaki1979systematic}.
In Fig.\ref{polyhedra} we plot the charge centres($\langle \phi |\vec{r}|\phi \rangle$)  of the approximate eigen states
of the first moment matrices. 
%
%
Evidently, the charge centres constitute coordination polyhedra around isolated atoms which are consistent in shape 
with those tabulated in Figs.6-8 in Ref.\cite{king2000atomic}.
This agreement confirms the identity of the resultant orbitals 
as the hybrid orbitals and numerically establishes the connection between maximal localization and hybridization.
Such a connection between $sp^3$ hybridization and  minimization of total quadratic spread of $s$ and the three $p$ orbitals
has been analytically proven\cite{nacbar2014simple}. 
In this work however we do not use GTOs further and rather resort to KS states of isolated atoms.
%
For example, for atoms of the $p$ block, such as boron, carbon, nitrogen and silicon dealt with in this work,
if the first moment matrices are constructed in the basis of three(four) KS states with lowest energies, namely, 
the one $s$ like non-degenerate having the lowest energy and two ( three) of
the three $p$ like degenerate states above the $s$ like state,
the approximate eigen subspace would render three(four) $2sp^2$($2sp^3$) hybridized orbitals.
%
%
Notably, for isolated systems like molecules, clusters and nanostructures, 
the approximate common eigen spectrum of the FMMs computed within the manifold of occupied KS states
results into partitioning\cite{singh2005novel,bhattacharjee2015activation,maji2019synergistic} 
of the ground state charge density into bonding and localized orbitals.
%
%

\subsubsection{Orientation and transfer of orbitals }
\begin{figure}[b]
\includegraphics[scale=0.27]{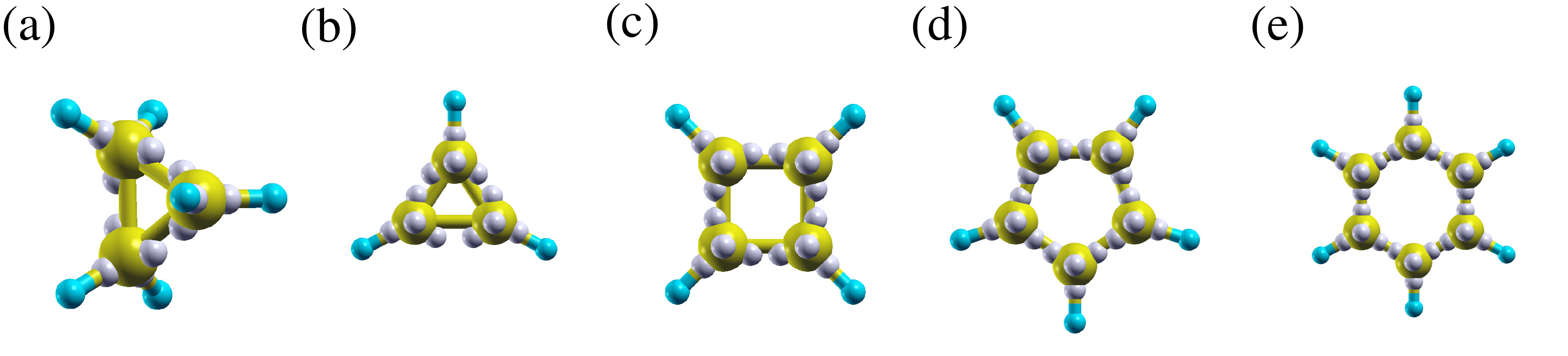}
\caption{Projected charge centres of HAOs are shown by gray spheres depicting their orientations 
around their host C atom shown in yellow.}
\label{molecules}
\end{figure}
\begin{figure}[t]
\includegraphics[scale=0.17]{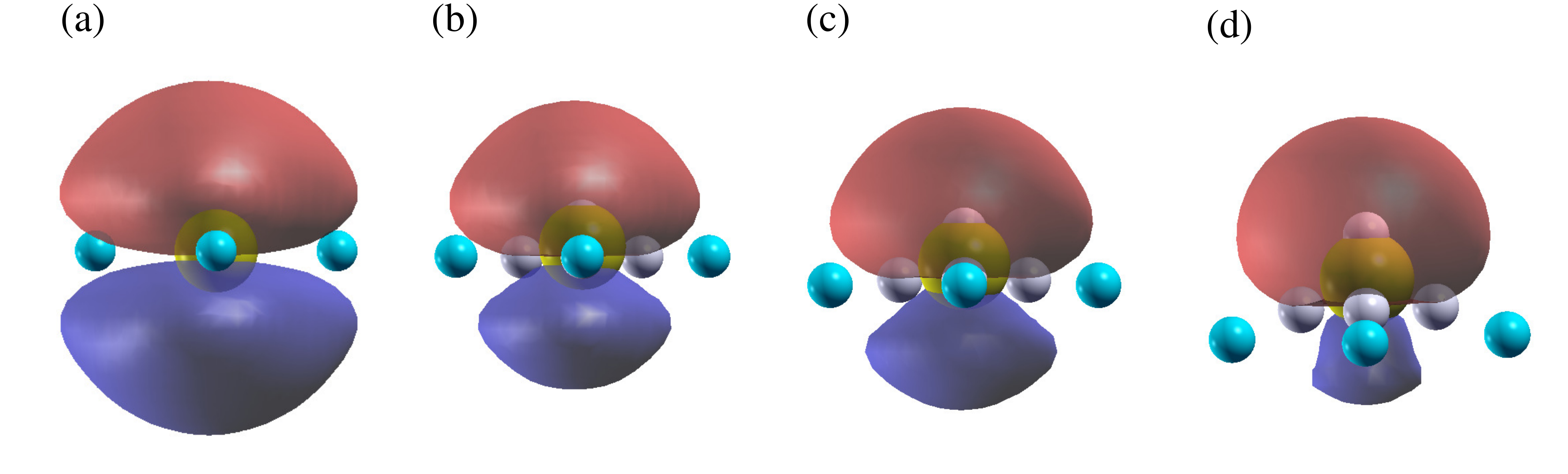}
\caption{(a-d): Evolution of a pure 2\pz \\ orbital[(a)] from \sptwo \\ hybridization background,
to an \spthree \\ hybridized orbital due to increased deviation of the centres (cyan spheres) 
of the three confining potential spheres from co-planarity with the host atom (yellow sphere). 
Centres of HAOs are shown by gray spheres. }
\label{scheme}
\end{figure}
%
\label{potref}
Although as evident above that construction of HAOs for an isolated atom as such do not require any pre-defined directionality,
the orientation of the HAOs associated with an atom can be nevertheless locked to their anticipated neighbourhood 
 by placing the isolated atom within an external potential
which represents the generic or exact atomic neighbourhood of the given atom in the actual system in which the HAOs are to be used. 
We construct such externals potentials by placing weakly confining spheres with small constant negative potentials inside the spheres
in place of exact or generic locations of neighbouring atoms as present in the actual system. 
For example, to lock \spthree \\ HAOs to a  four coordinated tetrahedral neighbourhood, 
a tetrahedra of confining spheres is placed around the host C atom,
leading to orientation of the \spthree \\ orbitals maximally in the direction of the confining spheres 
as seen in Fig.\ref{nhawo_ada}(a).
Typically we find confining potential amplitudes in the order of 0.01 eV and radius 0.5\AA 
to be sufficient for the purpose. 
Such weak confinement in the vicinity causes change of KS energy eigen-values of isolated atoms
 in the order of 0.001 eV, 
and retains the shape of the lowest KS states which are used for construction of the HAOs, 
effectively unaltered.
For \spthree\\ HAOs, 
the tetrahedra of the confining spheres can be an exact tetrahedra, as in case of bulk Si, or a strained tetrahedra,
as in case of  cyclopropane.
As evident in Fig.\ref{molecules}(a) for cyclopropane, 
and in Fig.\ref{molecules}(b-e) for planar molecules  C$_n$H$_n$, 
the projected charge center of the HAOs (shown in gray) symmetrically deviate away from the C-C bonds with 
decreasing C-C-C angle as we go from C$_6$H$_6$ to C$_3$H$_3$.
For all of these molecules the HAOs were constructed with the weakly confining spheres placed around the 
host C atom exactly as 
per their nearest neighbours in the molecules, resulting into HAOs largely retaining their pure $sp^3$ nature
but oriented symmetrically about the directions of the confining spheres from the host atoms.
The placement of confining potential spheres thus provide a gross directional reference for orientation of the 
full set of the HAOs.

\begin{figure}[t]
\includegraphics[scale=0.3]{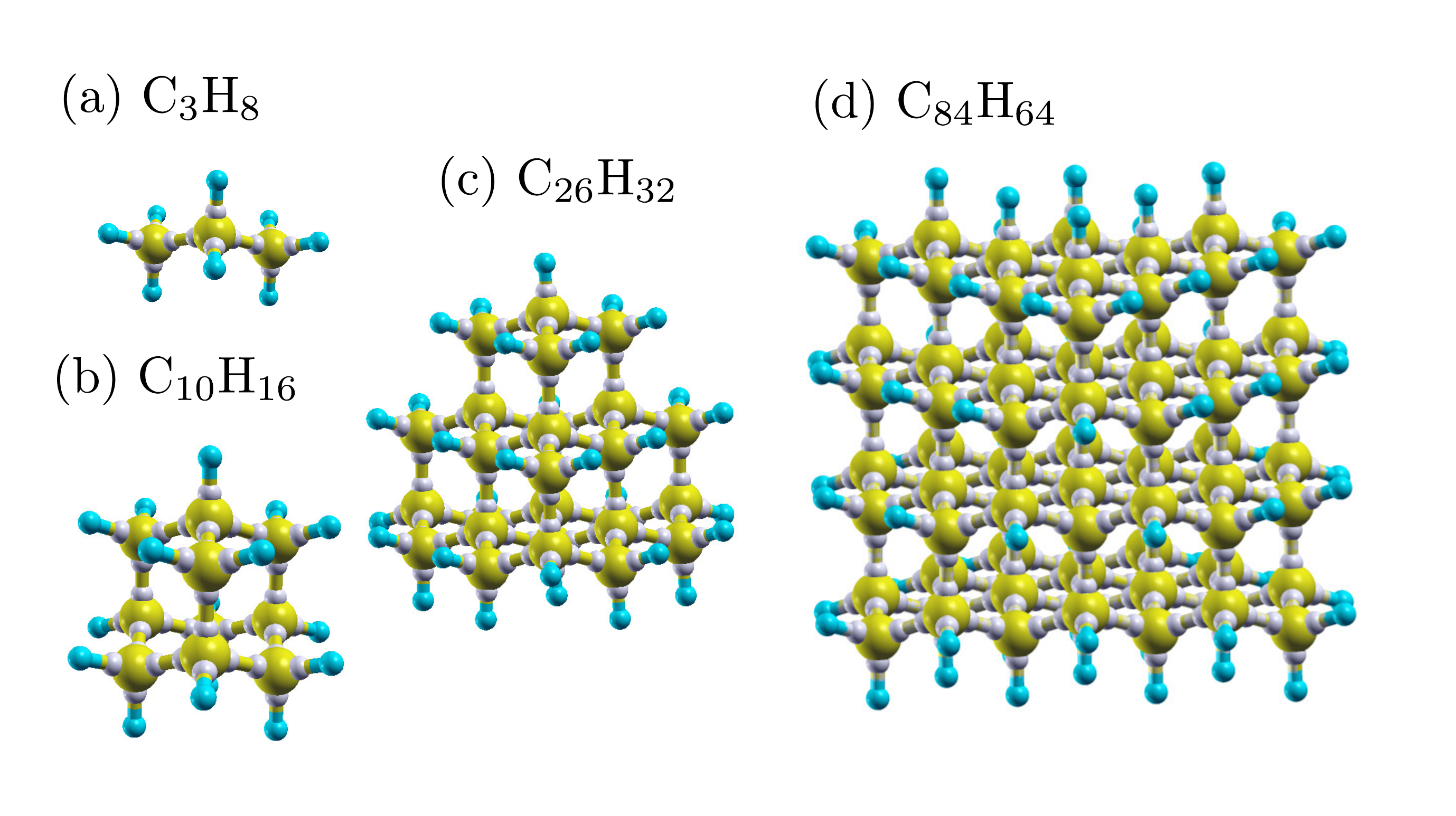}
\caption{C$_n$H$_m$ systems with projected charge centre of HAOs shown as gray spheres, used
in this work as example of \spthree\\ hybridized covalent systems.}
\label{wcmap}
\end{figure}

\begin{figure}[b]
\includegraphics[scale=0.25]{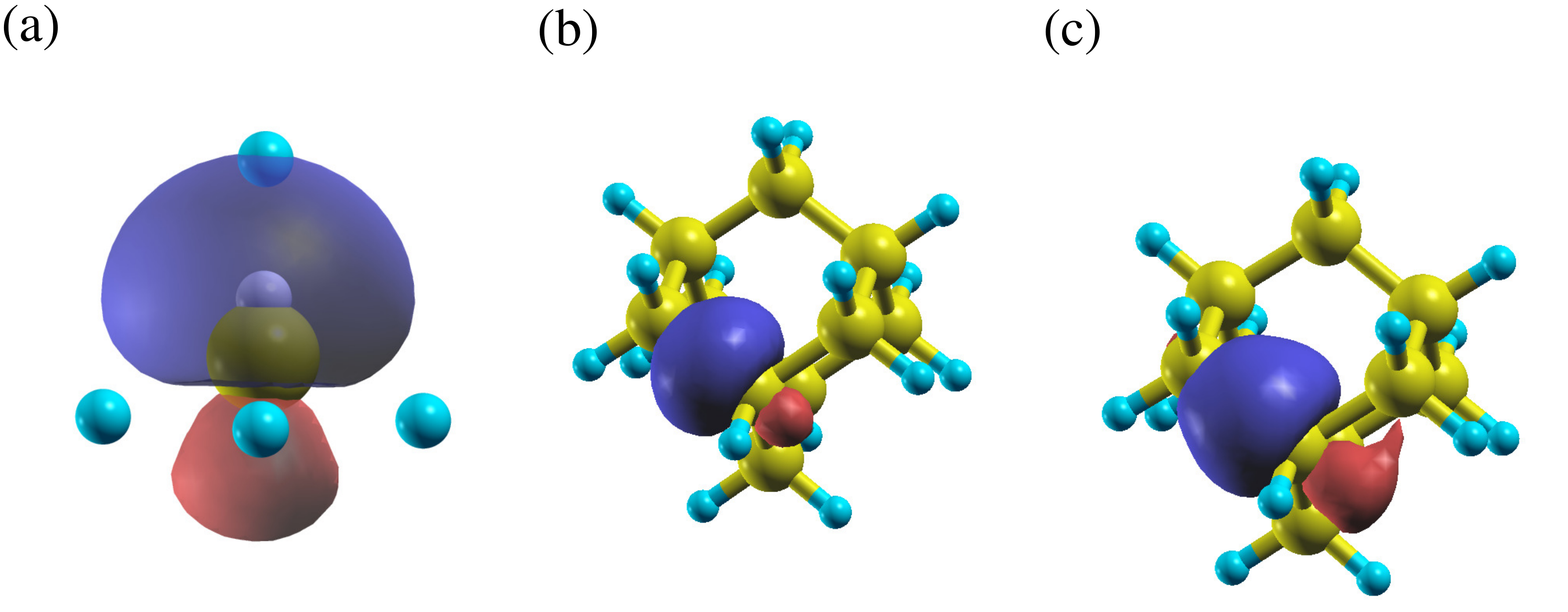}
\caption{(a): HAO representing a \spthree \\ orbital of an isolated C atom (yellow sphere) used in this work. 
Charge centre of the orbital is shown in gray. 
Centres of the  confining spheres used to determine gross orientation are shown in cyan.
(b): HAO shown in (a) transfered to a C atom an adamantane(\ada\\) molecule, 
(c): the corresponding HAWO.}
\label{nhawo_ada}
\end{figure}

Position of charge centres of the HAOs are learned in terms of the directions of the confining spheres 
from the isolated host atom.
%
%
Such learnings are subsequently used in projecting centres of HAOs around the 
corresponding atom in a given system, as seen for the molecules in Fig.\ref{molecules}, 
and nano-diamonds in Fig.\ref{wcmap}. 
While transferring HAOs from their nursery of isolated host atoms, to their matching host atoms in a given system,
HAOs are rotated such that their actual charge centres align along the direction of their projected centers 
from the matching host atoms.


In addition to providing reference for orientation, the confining spheres can have an important role 
in deciding the level of hybridization of the HAOs.
%
%
This becomes evident by noting that if we use four KS states and three confining sphere coplanar with the host atom,
then instead of forming four \spthree \\ orbitals,
 the HAOs separate into three 2\sptwo \\ orbitals and one 2\pz \\ orbital, as evident from the unhybridized shape 
of the 2\pz \\ orbital in  Fig.\ref{scheme}(a).
Fig.\ref{scheme}(a-d) shows evolution of the 2\pz \\ HAO from a pure orbital perpendicular to the plane 
of \sptwo \\ hybridization, towards a 2\spthree \\ hybridized orbital, with increasing non-coplanarity 
of the confining spheres with the host atom. 
HAOs with such intermediate hybridization  (2$sp^{2+}+2p_z^+$) has been used for \csixty \\[Fig.\ref{c60}].
However,  stronger confining potentials are found necessary to 
influence hybridization of KS states, typically in the order of 1eV  for C atoms, 
such that the orbitals align along the confining spheres.
The confining potentials in this case therefore does lead to minor modification of shape of the KS states,
and thereby of the HAOs as well, although not quite obvious at the iso-surfaces plotted  in Fig.\ref{scheme}(a-d).
However the  values of TB parameters calculated in the basis of their
Wannierized counterparts in \csixty\\ suggests that the overall shape of those orbitals are largely 
retained close to the \sptwo\\  orbitals.
Notably, we could have used stronger confinement to align the HAOs in C$_3$H$_6$, C$_3$H$_3$ or C$_4$H$_4$ as well like we did for \csixty\\,
but the degree of confinement would have to be much high than that used for \csixty\\, which would have substantially altered the shape of
the HAOs themselves, since it is obvious that with pure $s,p_x,p_y,p_z$ orbitals it is impossible to form
any set of hybrid orbitals in which two orbitals can have relative orientation less than 90$^\circ$.


%
%

%

\begin{figure}[b]
\includegraphics[scale=0.32]{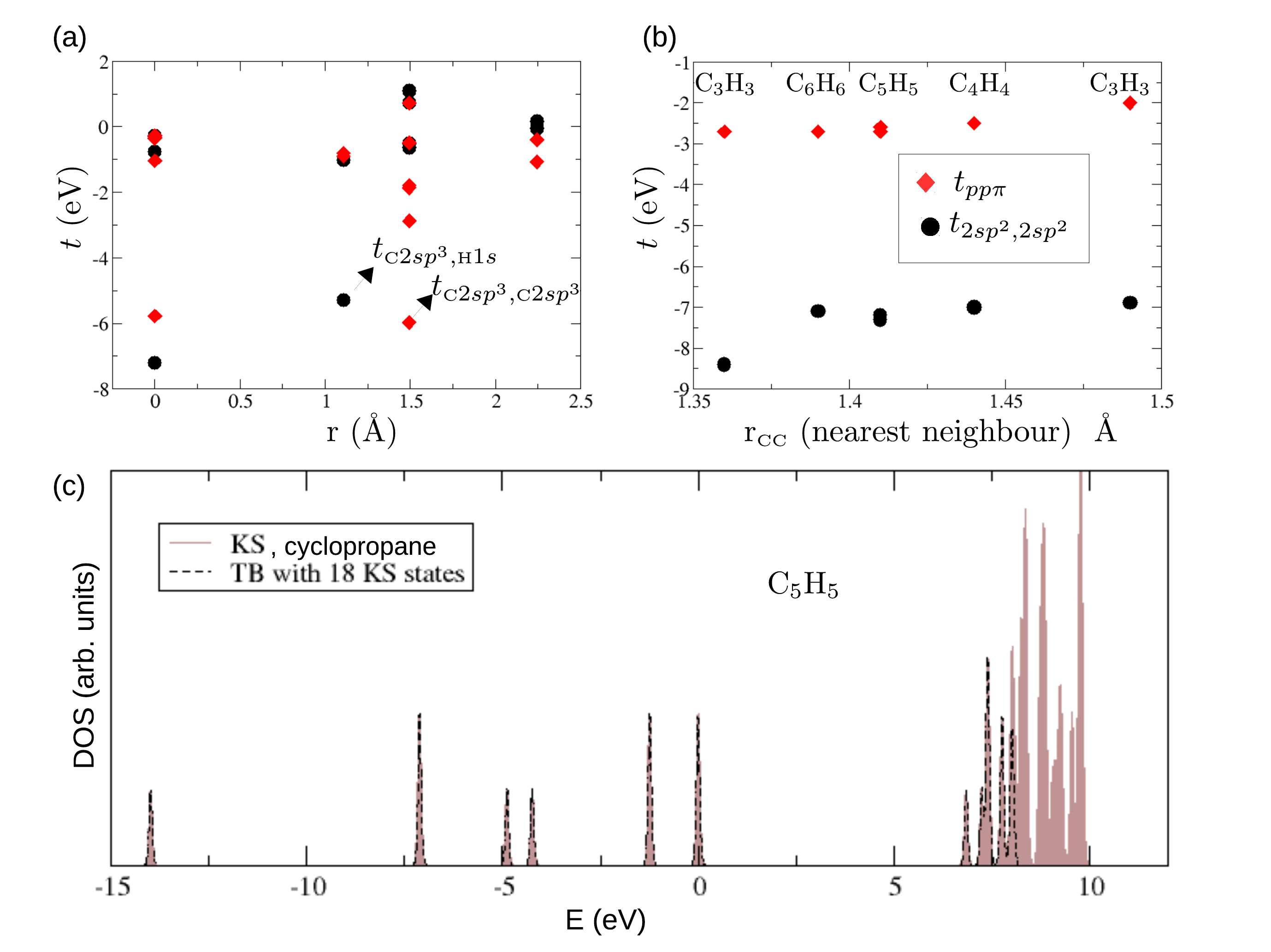}
\caption{(a): TB parameter calculated for cyclopropane; (b): Nearest neighbour TB parameters between in-plane and 
out of plane orbitals in C$_3$H$_3$, C$_4$H$_4$, C$_5$H$_5$ and C$_6$H$_6$ 
molecules (shown in Fig.\ref{molecules})
arranged as a function of C-C bond lengths available in the molecules.
(c) DOS calcaulated from 50 lowest KS eigen-values, compared with DOS from eigen-values of 
TB hamiltonian constructed from 18 lowest KS states, 
18 being the total number of valence orbitals of cyclopropane.}
\label{tb_molecules}
\end{figure}

\begin{figure*}[t]
\includegraphics[scale=0.33]{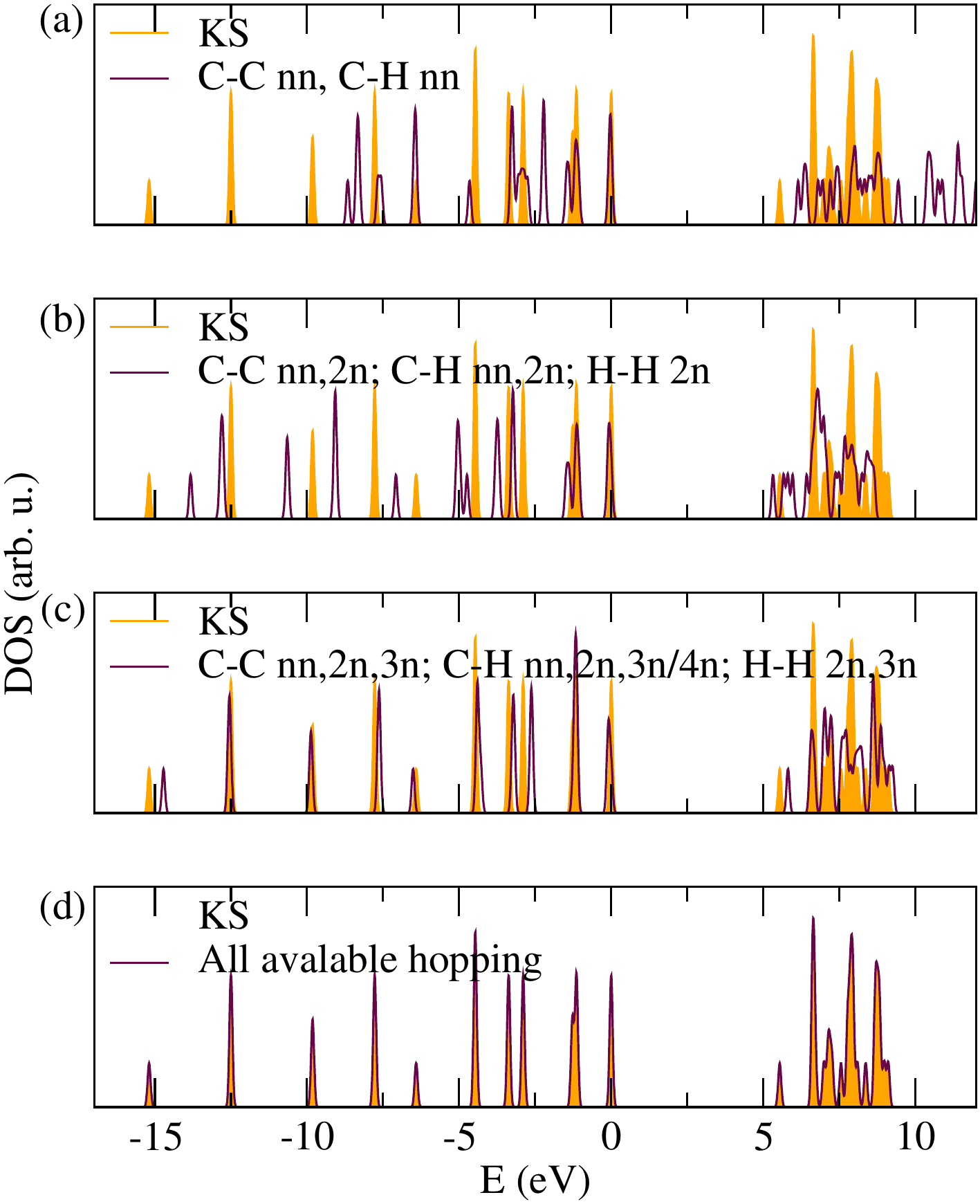}
\includegraphics[scale=0.35]{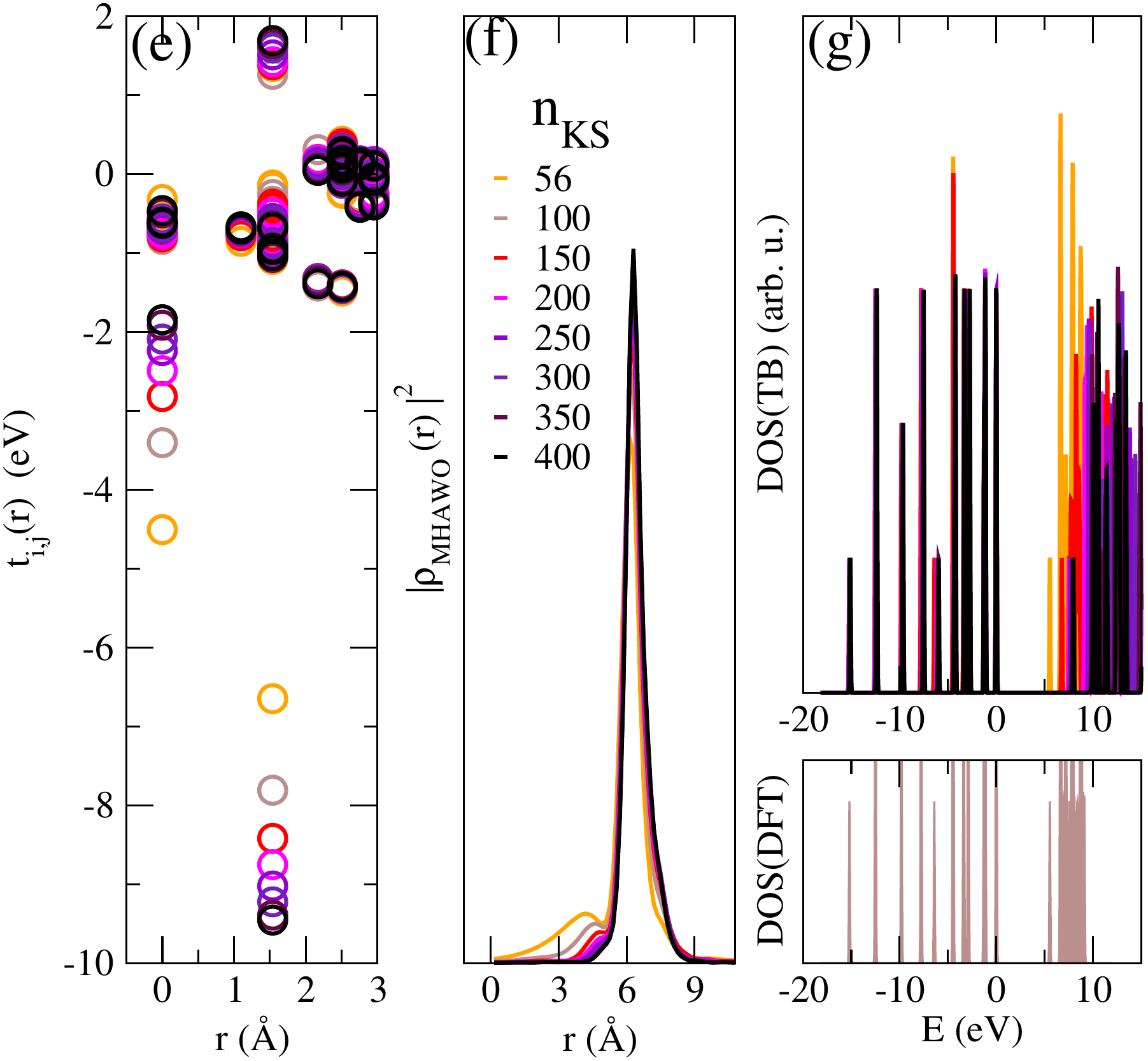}
\includegraphics[scale=0.35]{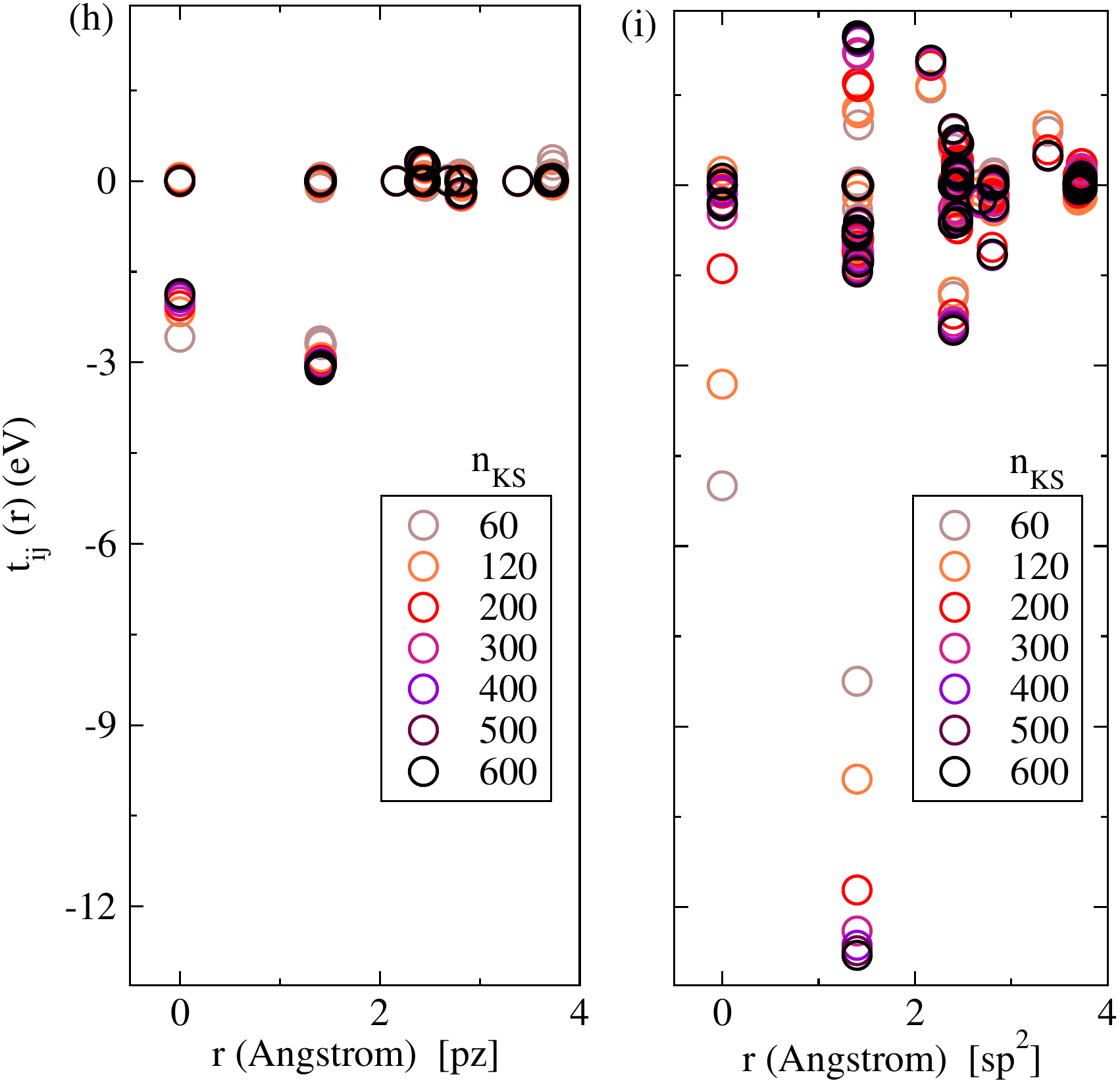}
\caption{For \ada \\, (a-d): Evolution of density of states(DOS) with increase in range of hopping starting from 
(a): the nearest neighbour (nn)
to (d): all avaliable hopping graduating through hopping between second (2n) and third (3n) nearest neighbours and beyond.
Convergence of (e): TB parameters and (f): spatial localization of 2\spthree \\ orbitals,
and (g): TB DOS, in terms of the number of KS states used  in construction of HAWOs as mentioned in the legend of (f).
KS DOS is shown below (g).
Similar convergence of TB parameters for (h): 2\pz\\ and (i): 2\sptwo\\ orbitals in AGNR (3p+1, p=2). }
\label{hopping}
\end{figure*}
%
\subsection{Wannier functions based on HAOs}
\label{mhawo}
The next step is to construct orthonormalized Wannier functions from the KS states following the HAOs transferred
to a given system. 
The transferred HAOs constitute a non-orthogonal basis of hybridized atomic orbitals.
In the general framework of periodic systems with non-zero wave-vectors ($\vec{k}$)
we begin with constructing a non-orthogonal set of quasi-Bloch states as:  
\begin{equation}
\tilde{\psi}_{\vec{k},j}(\vec{r}) = \frac{1}{\sqrt{N}} \sum_{\vec{R}} e^{i \vec{k} \cdot \vec{R}} \phi_{\vec{R},j} (\vec{r}),
\end{equation}
where $ \phi_{\vec{R},j} (\vec{r})$ is the $j$-th HAO localized
in the unit-cell denoted by the lattice vector $\vec{R}$ spanning over $N$ unit-cells defining the Born-von Karman periodicity.
The projections of the non-orthogonal quasi-Bloch states on the orthonormal Bloch states constructed from the 
KS single-particle states at all allowed $\vec{k}$, are calculated as:
\begin{equation}
 O_{\vec{k},m,j}= \langle \psi^{KS}_{\vec{k},m}  \mid \tilde{\psi}_{\vec{k},j} \rangle.
\label{proj}
\end{equation}   
Elements of $O$ thus record the representation of the HAOs within the manifold of KS bands considered.
Overlaps between the non-orthogonal quasi-Bloch states within the manifold of the considered KS states 
are therefore calculated as:
\begin{equation}
S_{\vec{k},m,n}= \sum_l O^*_{\vec{k},l,m} O_{\vec{k},l,n}. 
\label{overlap}
\end{equation}   
%
The degree of representability of HAO $\phi_n$, within the set of KS states considered, 
is guaranteed by setting a lower cutoff on individual $S_{\vec{k},n,n}$ values to be typically more than 0.85.
For all the system studied in this work, the above criteria is found to be satisfied by the lower bound on the number 
KS states, which is set by the total number of valence orbitals of all atoms of a given system.  
A new set of orthonormal Bloch states from the KS single particle states
are subsequently constructed using the L\"{o}wdin symmetric orthogonalization \cite{lowdin1950non} 
scheme as:
\begin{equation}
\Psi_{\vec{k},n}(\vec{r}) = \sum_m S^{-\frac{1}{2}}_{\vec{k},m,n} \sum_l O_{\vec{k},l,m} \psi^{KS}_{\vec{k},l}(\vec{r}),
\label{awobf}
\end{equation}
where the sum over $l$ spans the KS states considered and the sum over $m$ takes care of the orthonormalization.
Subsequently, a localized set of orthonormal Wannier functions are constructed as:
\begin{equation}
\Phi_{\vec{R'},j}(\vec{r}) = \frac{1}{\sqrt{N}} \sum_{\vec{k}}  e^{-i \vec{k}\cdot \vec{R'}} \Psi_{\vec{k},j}(\vec{r}).
\label{awo}
\end{equation}   
%
In this process the L\"{o}wdin symmetric orthogonalization clearly provides a  choice of gauge for 
linear combination of KS states such that the resultant Wannier functions $\left\{ \Phi_{\vec{R'},j}(\vec{r}) \right\}$ 
resemble the corresponding HAOs $[\left\{ \phi_{\vec{R},j} (\vec{r})  \right\}]$ as much possible within the manifold 
of KS states considered.
Hence we refer to these Wannier functions as the hybrid atomic Wannier orbitals (``HAWO''). 
In Fig.\ref{nhawo_ada}
we show an HAO before and after transfer to adamantane and the corresponding HAWO constructed from the KS states of adamantane.
%
%
HAWOs can thus be considered as analogue of NHOs constructed from a given set of KS states with acceptable representability. 
%
%
\subsubsection{TB parameters in HAWO basis}
TB parameters in the HAWO basis are computed from energetics of KS single particle states as:
\begin{eqnarray}
& &t_{\vec{R'},\vec{R},i,j} = \langle \Phi_{\vec{R'},i} \mid H^{KS} \mid \Phi_{\vec{R},j} \rangle \nonumber \\
&=&\sum_{\vec{k}}e^{i\vec{k}.(\vec{R'}-\vec{R})}\sum_l (OS^{-\frac{1}{2}})^*_{li} (OS^{-\frac{1}{2}})_{lj} E^{KS}_{\vec{k},l}
\label{hop}
\end{eqnarray}
Notably, similar TB parameters have been derived in the last two decade from first principles based on the either the
maximally localized Wannier function \cite{marzari2012maximally,lee2005band,atomicorb2,calzolari2004ab,franchini2012maximally,jung2013tight} 
or atomic orbitals \cite{atomicorb1,qian2010calculating} constructed from KS states.
Much effort has been reported in deriving TB parameters through projection of KS states on 
pseudo-atomic orbitals \cite{d2016accurate,agapito2016accurate} as well.
%
%
However, attempts to calculate TB parameters in hybrid atomic orbital basis constructed from first-principles,
as proposed in this work, has been limited so far primarily to analytical models\cite{yue2017thermal,popov2019deductive}. 
%

%
In Fig.\ref{tb_molecules}(a)  
for cyclopropane, we plot the TB parameters calculated as per Eqn.(\ref{hop}) 
 for two HAOs participating dominantly in a C-C bond and a C-H bond. 
The $t_{sp^3,sp^3}$ is comparable to the that in adamantane (\ada \\)[Fig.\ref{SEChopping}] despite the substantial misalignment[Fig.\ref{molecules}] of HAO 
and the C-C bond in cyclopropane while perfect alignment of the two in \ada\\. 
The hopping parameters are obtained with 18 KS states which is same as the total number of valence orbitals
of all the atoms,
resulting thereby into density of states in exact agreement with that obtained from DFT [Fig.\ref{tb_molecules}(c)]
as discussed above in the next paragraph.
In Fig . \ref{tb_molecules}(b)
we plot hopping parameters for $\pi$ and $\sigma$ bonds as a function of C-C bond lengths available in 
planar C$_3$H$_3$ to C$_6$H$_6$ molecules.
As evident in Fig.\ref{molecules},  
the best alignment of the HAOs along the C-C bond is possible for C$_6$H$_6$ and the worst is obviously for the shorter bond
of C$_3$H$_3$ and similarly for C$_3$H$_6$.
Yet, the highest in-plane hopping parameter in terms of magnitude is found for the shorter bond of C$_3$H$_3$, which is 
about 20 \% more than that of the  C-C in-plane bond of benzene,
whereas the  C-C bond length in benzene only about 2.2 \% more than the  shorter bond of C$_3$H$_3$.   
Similarly, the C-C nearest neighbour hoping parameter as well as the bond length in C$_3$H$_8$, both are within 1 \% of 
those of  C$_3$H$_6$, whereas in C$_3$H$_8$ the HAOs are almost perfectly aligned along the C-C bond [Fig.\ref{wcmap}] 
while in C$_3$H$_6$ they are misaligned by more than 20$^\circ$.
These results can possibly be explained by the inherent bent nature the bonds\cite{wiberg1996bent} 
in C$_3$H$_6$ and  C$_3$H$_3$, reflected by the symmetric misalignment of the HAOs along the two C-C bonds
while maintaining perfect alignment along the C-H bonds. 
We plan to examine this aspect for bent bonds in details in future. 

%
As evident in  Fig.\ref{hopping}(a) for \ada, the edge of the valence band is already well described if we consider
only the nearest neighbour hopping in the HAWO basis.
However, as shown in  Fig.\ref{hopping}(b) onwards, the match of DOS from TB and DFT improves drastically with 
increasing extent of hopping considered up to the second nearest neigbour. 
This is immediately understood by noting the non-nominal positive valued of the second nearest hopping element 
plotted in figure Fig.\ref{hopping}(e), arising due to proximity of lobes of different signs of the two HAOs.
In  Fig.\ref{hopping}(e-g), we demonstrate evolution  of the TB parameters, HAWOs, and DOS from TB, 
as function of number of KS states considered for construction of HAWOs.
The rationale for this analysis is the possibility that the anti-bonding subspace may not be adequately represented by
the unoccupied KS states if we restrict the total number of KS states to be same as the total number of HAOs associated with all the atoms, which is same as the total number of valence orbitals of all the atoms.
Indeed we see clear convergence of shape of  HAWO [Fig.\ref{hopping}(f) ]
as well as the corresponding TB parameters[Fig.\ref{hopping}(e)] 
if we consider KS states in excess of the total number of HAOs.
Fig.\ref{hopping}(h,i) suggests that the convergence can be much quicker for un-hybridized orbitals like 2\pz\\,  
compared to  hybridized orbitals like \sptwo\\ and \spthree \\, since the un-hybridized orbitals primarily constitutes
the edges of the valence and conduction bands.
However, the TB DOS expectedly starts deviating from the DFT DOS more in the conduction band [Fig.Fig.\ref{hopping}(g)] 
 if we include more KS states beyond 
the total number of HAOs, owing to the semi-unitary nature of the net transformation 
matrix $(OS^{\frac{1}{2}})$ implied in Eqn.(\ref{awobf}) which will be rectangular in such scenarios.
It is thus important to decide on the number of KS states to be considered depending on the purpose.
If the aim is to represent only the valence bands through well localized HAWOs, then it may be prudent to look for convergence 
of HAWOs in terms of KS states. But if band-gap needs to be represented accurately by the TB parameters then the number of KS states 
should be kept same as the total number of valence orbitals.
%

\subsection{Bottom-up mapping of TB parameters}
\label{mapr2t}
The HAWO basis derived from the KS states offer a multi-orbital TB basis which are by construction 
locked to the local coordination as per the atomic neighbourhood of each atom.
The TB parameters derive in such a basis should therefore be transferable from one system
to another with matching atomic environment.
A key aim of this work is to demonstrate such transferability for effective transfer
of multi-orbital TB parameters in the HAWO basis from smaller reference systems to larger target systems.
%
%
%
The mapping of TB parameters is done in two steps.\\ 
(1) Pairs of atoms of the target system, not limited to nearest neighbours,
are mapped on to pairs of atoms in the reference system based on a collection of criteria.\\
(2) Among the mapped pair of atoms, pair of system orbitals are mapped to pair of reference orbitals 
through mapping of their respective projected charge centres. 
%
%
In step (1) the criteria to map pairs of atoms include matching structural parameters such as 
their spatial separation and their individual nearest neighbourhoods characterised in terms of the type of neighbouring atoms 
and angles made by nearest neighbours on the atoms.
In particular, we use a  parameter calculated as:
\begin{equation}
\zeta_i = \sum^{N_i}_j Z_j w(r_{i,j})
\label{mapparam}
\end{equation}   
where $N_i$ is the number of neighbours of the $i$-th atom within a suitably chosen cutoff radius, 
$w$ being a weight factor which is a function of the distance $r_{i,j}$ of the $j$-th neighbour of the $i$-th atom,
and $Z_i$ a characteristic number to be associated with each type of atom.
$Z_i$ can be chosen to be the atomic weight, as we mostly used in this work, or a similar number which can facilitate
identification of a type of neighbourhood or a region of the system through values of $\zeta$.
In this work we chose the weight factor $w$ to be 1.0 within half of the cutoff radius beyond which the factor
is smoothly reduced to zero using a cosine function.  
The choice of cutoff radius depends on the size of the reference system. 
It should neither be too large for variations to average out, nor should it be too small to become insensitive to
morphological variations in the reference system itself. 
%
%
$\zeta$ allows us to map atom pairs effectively through prudent choice of values of $\left\{ Z_i \right\}$ 
since it would allow assessment of proximity of atoms to edges, interfaces or any kind of structural inhomogeneity 
without any exhaustive structural relaxation.

%
In step 1, the minimum of the deviation:
\[
|\zeta_1^{\mbox{target}}-\zeta_1^{\mbox{reference}}|+|\zeta_2^{\mbox{target}}-\zeta_2^{\mbox{reference}}|
\]
obtained within a range of allowed deviation of structural parameters, is used as the criteria to choose
matching pairs of atoms between target and reference systems.  

Like in step 1, in step 2 as well, the mapping of one or a pair of HAOs from the reference to target systems is done
based on matching structural parameters, as well as  a  parameter calculated as:
\begin{equation}
\xi_i = \sum^{N_i}_j \zeta_j w^{WC}(r_{i,j})
\label{mapwc}
\end{equation}   
where $\zeta_j$ corresponds of the $j$-th atom in the neighbourhood defined by $w^{WC}$ around 
of the projected charge centre of the $i$-the  HAO.
Angle made by the directions of the projected charge centres of the HAOs from their respective host atoms is a key matching
parameter in step 2.
Additionally, if the  HAOs  belong to different atoms then the dihedral angle made 
by the centres of the HAOs through the axis connecting their host atoms, is also a key parameter.
%
%
Thus in step 2, the minimum of the deviation  
\[
|\xi_1^{\mbox{target}}-\xi_1^{\mbox{reference}}|+| \xi_2^{\mbox{target}}- \xi_2^{\mbox{reference}}|
\]
within acceptable deviations of structural parameters, defines matching pairs of HAOs.

\begin{figure}[t]
\includegraphics[scale=0.32]{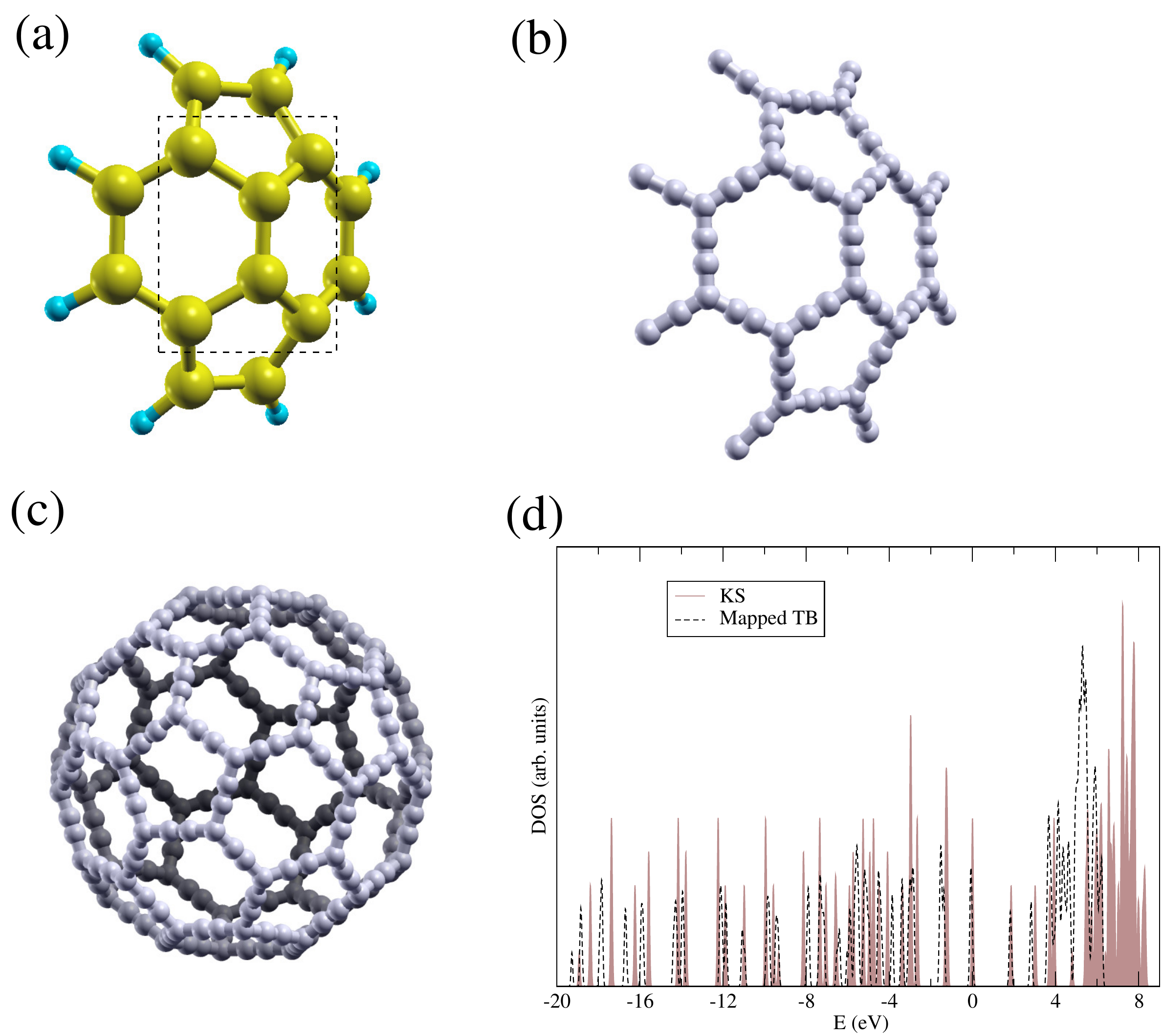}
\caption{(a): Structure of reference system, and (b): the corresponding charge centres of HAOs 
with intermediate hybridization (2$sp^{2+}$+2$p_z^{+}$) between \sptwo \\ and \spthree \\. 
(c): Projected charge centre with similar hybridization for \csixty \\.
(d): Corresponding matches of DFT DOS with TB DOS with parameters mapped from the reference system.
}
\label{c60}
\end{figure}

As an example we show mapping from a small curved finite patch[Fig.\ref{c60}(a)] to \csixty \\. 
Since \csixty \\ constitutes a curved surface without any edge, mapping should be done
from the inner most neighbourhood of the chunk.
Since in \csixty \\, the angles made by nearest neighbours at a given atom differ distinctly depending on whether 
an angle opens inside a pentagon or a hexagon,
the matching parameters for mapping are mostly structural, primarily the direct and dihedral angles.
The reference patch is cropped from \csixty \\ and passivated by H. 
We fix $\zeta$ tolerance to zero which implies that \csixty \\ is getting mapped from only six C atoms of the patch [Fig.\ref{c60}(a)]
having all C neighbours.
Given the curvature of \csixty \\, we chose to use confining spheres to influence the hybridization of \sptwo\\ HAOs
in order to break their coplanarity and align them along nearest neighbour C-C bonds,  as shown in Fig.\ref{scheme}(c)
where the placement of confining potential spheres are as per the nearest neighbourhood in \csixty \\.
The projected charge centres of HAOs with intermediate hybridization (2$sp^{2+}+2p_z^+$) between  (\sptwo\\+\pz\\) and \spthree \\ 
shown in Fig.\ref{c60}(c), is used to map  from that of the reference shown in Fig.\ref{c60}(b). 
TB parameters $t_{2sp^{2+},2sp^{2+}}$ for the shorter and longer C-C bonds are about -6.9 eV and -6.5 eV,
whereas  $t_{2p_z^{+},2p_z^{+}}$ are about -2.36 eV and -2.0eV. 
The match of the  DFT DOS with the DOS from TB parameters mapped from the reference system is shown in Fig.\ref{c60}(d),
which can be further improved beyond the valence bond by considering HAOs for excited states, which will be
taken up in a subsequent work on optical properties.
%

\begin{figure}[t]
\includegraphics[scale=0.45]{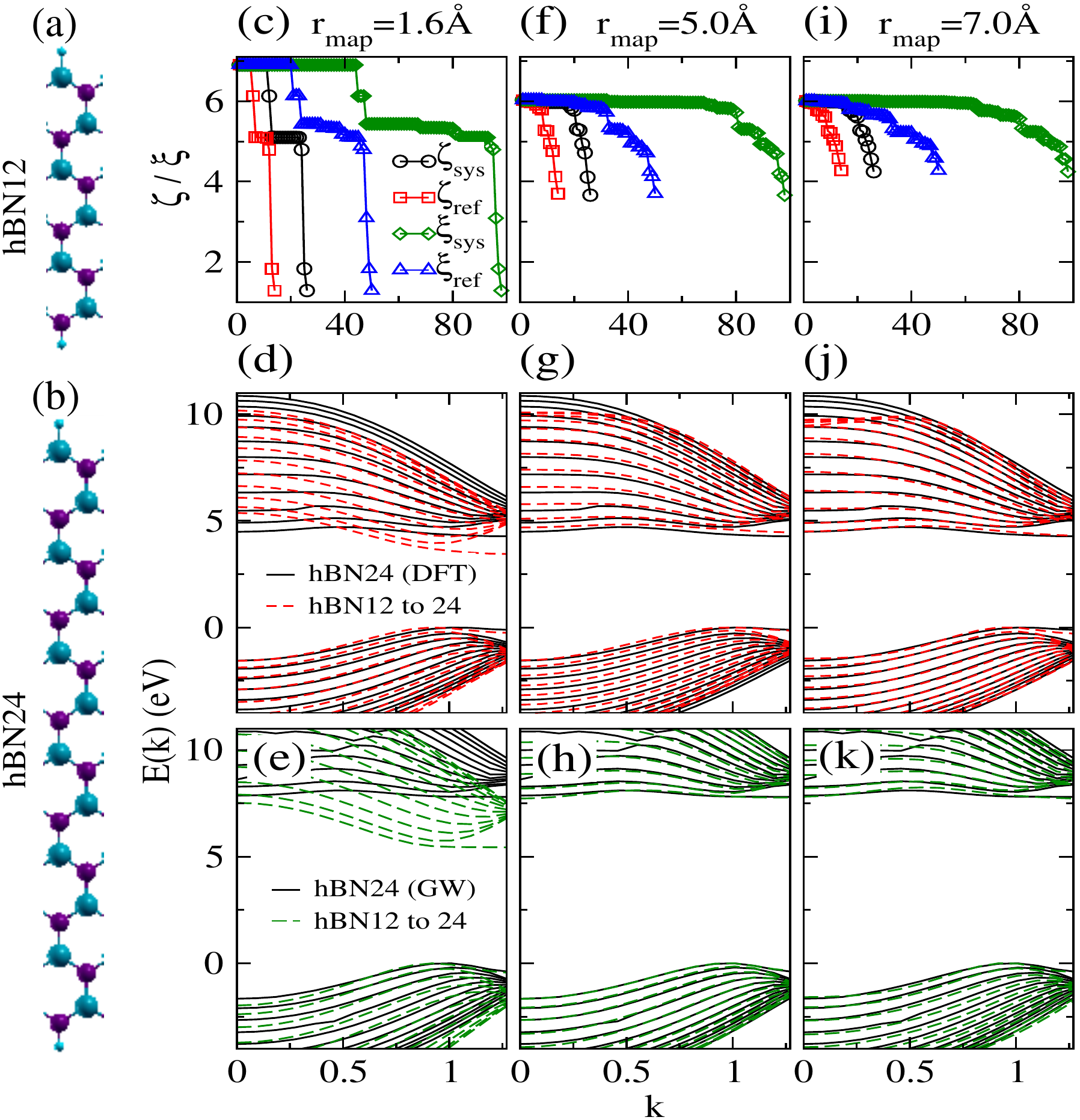}
\caption{(a,b) Hexagonal zigzag boron nitride nanoribbons(hZBNNR: hBN), hBN12 and hBN24 respectively. (c,f,i) Plot of  $\zeta$ and $\xi$ values ``ref''(reference hBN12) and 
``sys''(target hBN24) for different spatial ranges of  
neighbourhood considered for mapping. 
(d,g,j) Matching of DFT band-structure and mapped TB band-structure  for increasing $r_{map}$.
(e,h,k) Matching of DFT+G$_0$W$_0$ band-structure and mapped self-energy corrected TB band-structure  for increasing $r_{map}$.}
\label{hbn}
\end{figure}
\begin{figure}[t]
\includegraphics[scale=0.36]{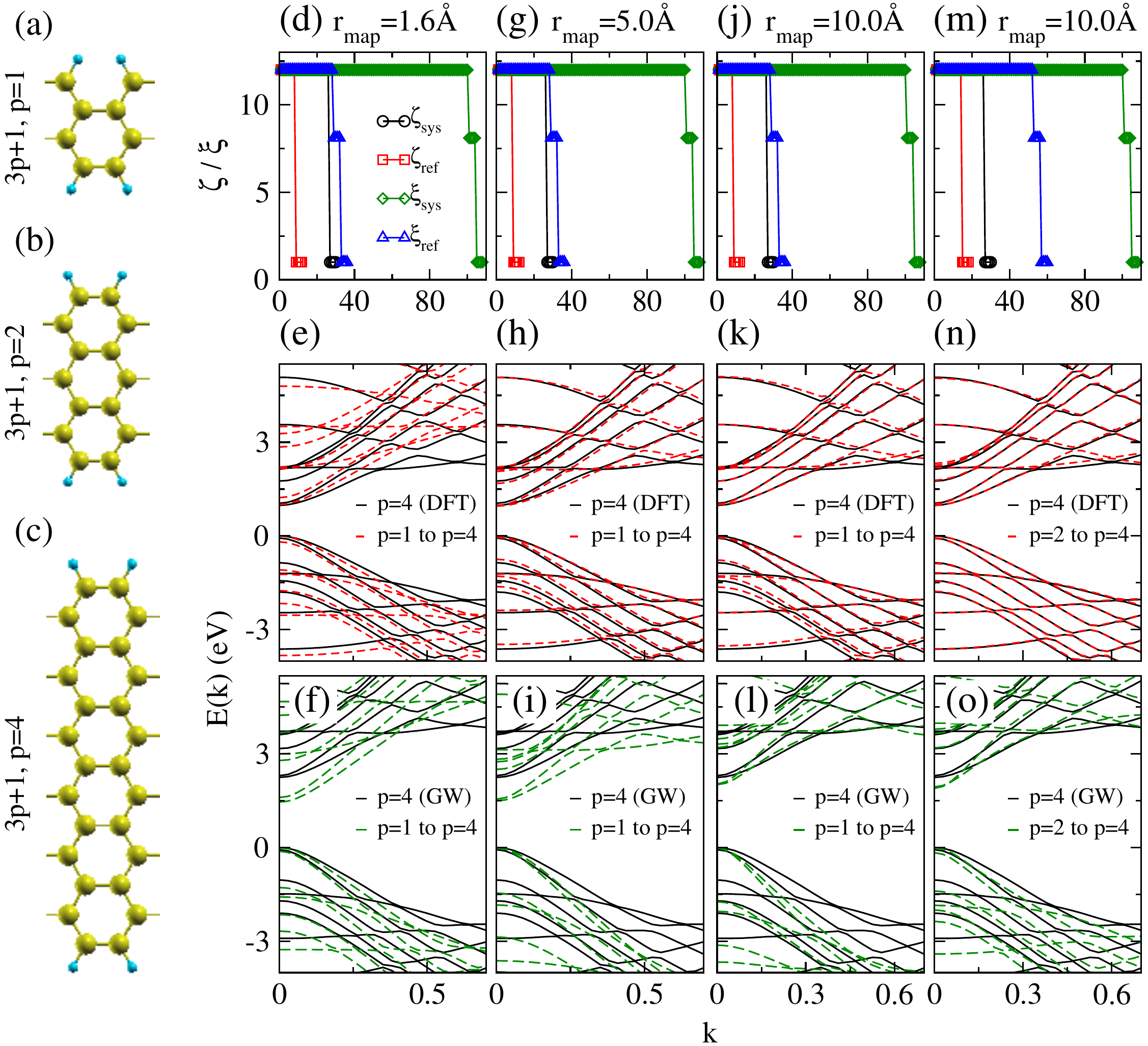}
\caption{(a,b,c) Armchair graphene nanoribbons (AGNR) of family n=3p+1 with p=1, p=2 and p=4 respectively. 
(d,g,j,m) Plot of  $\zeta$ and $\xi$ values ``ref''(reference p=1 (in d,g,j) 
and p=2 (in m)) and ``sys''(target p=4) for different spatial ranges 
($r_{map}$) of neighbourhood considered for mapping. 
(e,h,k,n) Matching of DFT band-structure and mapped TB band-structure for increasing $r_{map}$.
(f,i,l,o) Matching of DFT+G$_0$W$_0$ band-structure and mapped self-energy corrected TB band-structure  for increasing $r_{map}$.
}
\label{agnr}
\end{figure}
\begin{figure}[t]
\includegraphics[scale=0.45]{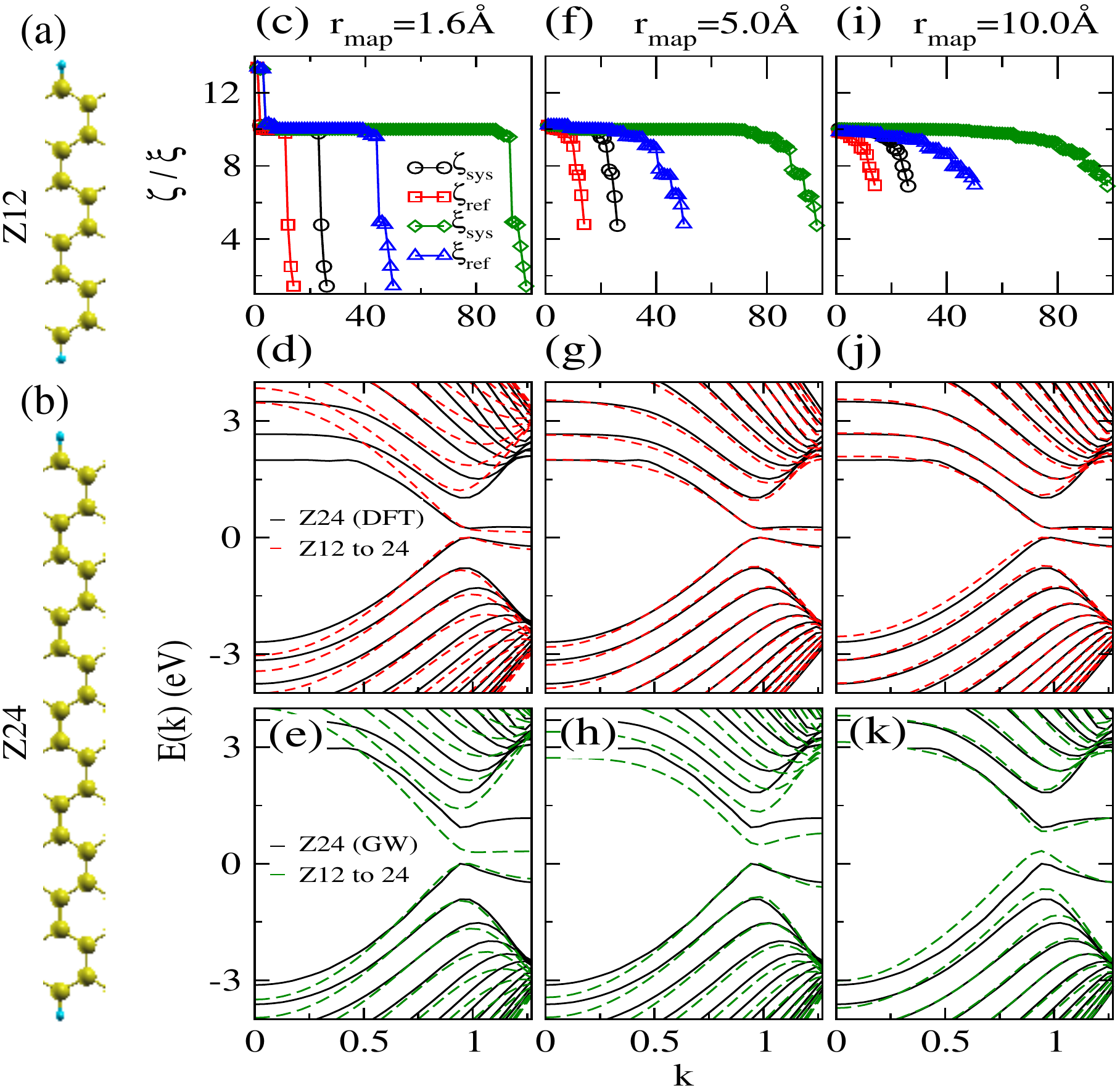}
\caption{(a,b) Zigzag graphene nanoribbons (ZGNR: Z), Z12 and Z24 respectively. 
(c,f,i) Plot of  $\zeta$ and $\xi$ values ``ref''(reference Z12) and 
``sys''(target Z24) for different spatial ranges of  
neighbourhood considered for mapping. 
(d,g,j) Matching of DFT band-structure and mapped TB band-structure  for increasing $r_{map}$.
(e,h,k) Matching of DFT+G$_0$W$_0$ band-structure and mapped self-energy corrected TB band-structure  for increasing $r_{map}$.
}
\label{zgnr}
\end{figure}


\subsection{Self-energy correction of TB parameters}

Self-energy corrected TB parameters $\left\{t^{QP}_{\vec{R'},\vec{R},i,j} \right\}$ in the HAWO basis are calculated 
by substituting $E^{KS}_{\vec{k},n}$ in Eqn.(\ref{hop}) by quasiparticle energies $E^{QP}_{\vec{k},n}$ obtained at
the $G_0W_0$ level which is the first order non-self-consistent GW approximation of MBPT\cite{hedin1965new} \cite{hedin1970effects}.
Within the GW approximation, the quasi-particle energies are approximated as:
\begin{equation}
E^{QP}_{\vec{k},n}=E^{KS}_{\vec{k},n}+\langle \psi^{KS}_{\vec{k},n} \mid \Sigma - V^{KS}_{xc} \mid  \psi^{KS}_{\vec{k},n} \rangle,
\label{equasi}
\end{equation}
where $ V^{KS}_{xc}$ 
is the mean-field exchange-correlation potential 
and  $\Sigma$\cite{hybertsen1986electron} is the self-energy operator 
derived by considering the many-electron effects as perturbation treated 
within a self-consistent framework of Dyson's equation 
formulated in terms of the one-particle dynamic non-local  Green's function
constructed from the KS states. 
%
Similar efforts have been reported in recent years on incorporating SEC in TB parameters 
computed in terms of the MLWFs\cite{qptb3,qptb5,qptb6}.
Incorporation of SEC in TB parameters has also been attempted 
through matching specific bands of the QP structure\cite{qptb1,qptb4,qptb7}.
%
\section{Computational Details}
Electronic structures of the ground states of all the systems considered in this work are calculated using the Quantum 
Espresso (QE) code\cite{giannozzi2009quantum} which is a plane wave based implementation of DFT.
We have used norm conserving pseudo-potentials with the Perdew-Zunger (LDA) exchange-correlation\cite{lda1}
functional and a plane wave cutoff of 60 Rydberg for wave-functions and commensurately more for charge density and potential.
Variable cell structural relaxation has been carried out for all periodic systems. 
%
We used a 1x1x15 Monkhorst-Pack grid of k-point for AGNRs and 1x1x29 for ZGNRs as well as for ZBNNRs.
Self-energy correction to single particle levels have been estimated at the non-self-consistent G$_0$W$_0$ level of GW approximation
implemented in the BerkeleyGW code\cite{deslippe2012berkeleygw}. 
To calculate the static dielectric matrix required for computation of the self-energy operator,
the generalized plasmon-pole model\cite{hybertsen1986electron} is used to extend the static dielectric matrix in the finite 
frequencies. 
For all the nanoribbons parameters are chosen from  Ref.\cite{yang2007quasiparticle}.
In house implementation interfaced with the QE code is used for generation of HAOs, HAWOs from KS states,
calculation of TB parameters in the HAWO basis, and mapping of TB parameters from reference to target systems.  
%
\subsection{Mapping self-energy corrected TB parameters in HAWO basis}
\label{mapping}
\begin{figure}[b]
\includegraphics[scale=0.33]{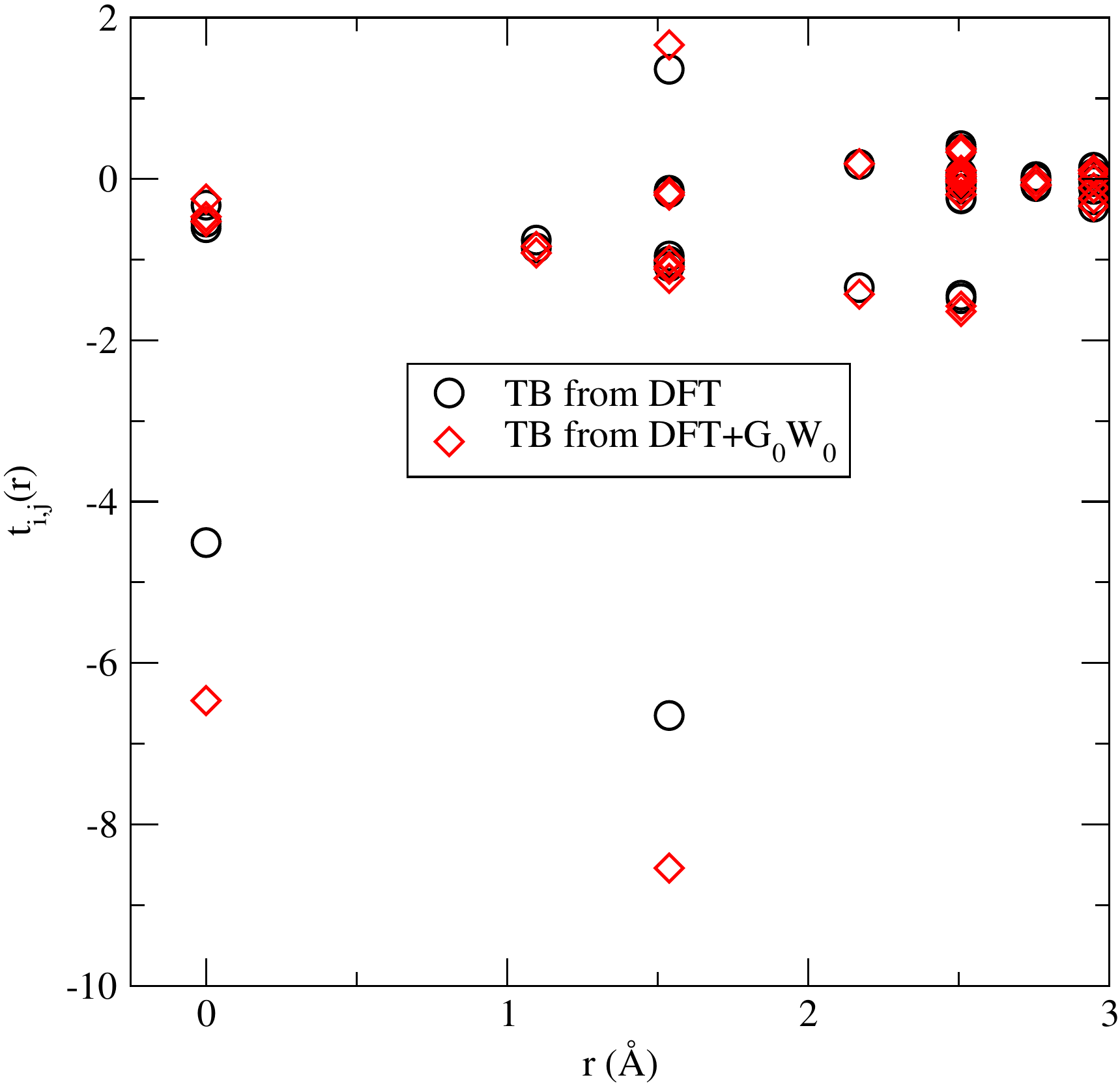}
\caption{TB parameters involving a C atom in \ada \\ with three C neighbours, 
computed using 56 KS states with and without SEC at the G$_0$W$_0$ level,
and plotted as a function of distance from the atom. 
TB parameters from DFT are same as those plotted in Fig.\ref{hopping}(e).}
\label{SEChopping}
\end{figure}
\subsubsection{Nanoribbons}
We have recently reported\cite{hossain2020transferability} estimation of 
quasi-particle(QP) band-gap for graphene and hBN nano-ribbons based on
SE correction to TB parameters 
mapped from  narrower ribbons in the basis of a single 2\pz \\ electron per atom. 
The transfer was made explicitly by identifying equivalent atoms based on proximity to the ribbon edges as shown in 
Fig.2(c-e), Fig.3(f) and Fig.5(a) in Ref\cite{hossain2020transferability}.
In this section we begin by systematizing the process of identifying the equivalent atoms through the mapping mechanism
proposed in Sec.\ref{mapr2t}. 
The identification is primarily based on $\zeta$ values for atoms and $\xi$ values for 
HAO charge centres wherever sufficient variations of $\zeta$ and $\xi$ are available in the reference systems, 
as demonstrated in Fig.\ref{hbn}(c,f,i) and Fig.\ref{zgnr}(c,f,i) for
hBN and ZGNR respectively.
Whereas, mapping of AGNR from p=1 to p=4, as shown in Fig.\ref{agnr}(d,g,j,m), 
calls for matching of structural parameters as the key strategy for mapping,
since the width of reference AGNR with p=1 of the 3p+1 family, is narrow enough and have only two types of C atoms per unit-cell.
%
\begin{figure}[b]
\includegraphics[scale=0.27]{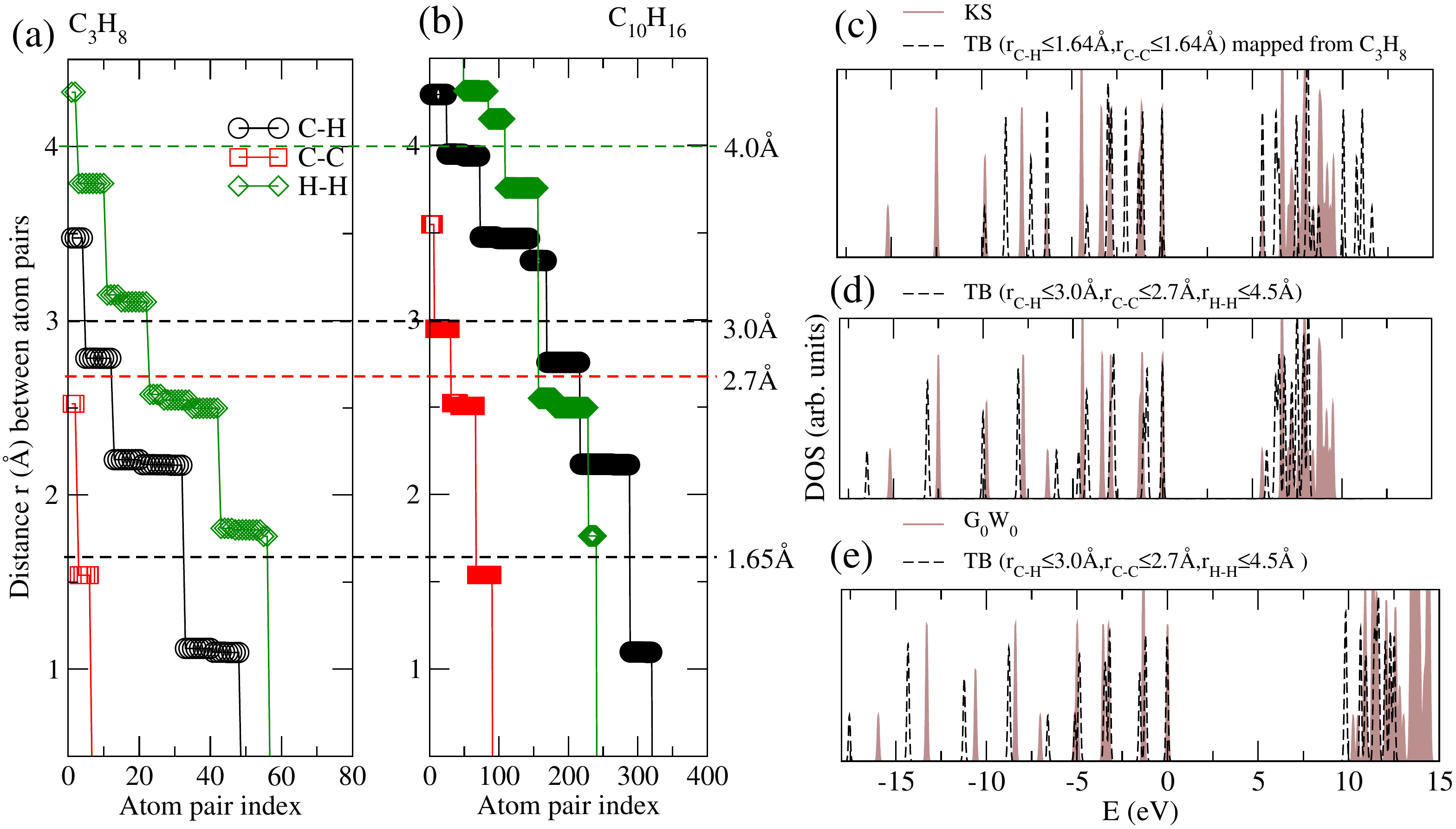}
\caption{Distribution of distance between pairs of atoms in (a): reference (\nds\\) and (b): target (\ada\\) systems. 
(c-d) Match between DFT DOS and mapped TB DOS as demonstrated of efficacy of mapping of TB parameters from 
\nds \\ to \ada \\ with  increasing spatial range of neighbourhood considered for mapping. 
(e)  Match between DFT+G$_0$W$_0$ DOS and mapped self-energy corrected TB DOS.}
\label{adamap}
\end{figure}

For hBN, acceptable match[Fig.\ref{hbn}(g,j and h,k)] 
between explicitly computed band-structure, and the same computed from TB parameters with only 2\pz \\ orbitals mapped from a 
narrower ribbon [Fig.\ref{hbn}(a)], is achieved simultaneously at the  DFT and DFT+G$_0$W$_0$ levels, 
with hopping considered within the range no less than  5\AA.
%
Whereas a higher spatial range of mapping of self-energy corrected TB parameters is required for matching of 
band-structure at the DFT+G$_0$W$_0$ level for AGNR
[Fig.\ref{agnr}(l)], and more so for ZGNRs [Fig.\ref{zgnr}(k)] with smaller band-gaps.
Notable for AGNR, the match of self-energy corrected band edges for p=4 naturally improve with mapping from p=2 [Fig.\ref{agnr}(o)]. 
These trends simply relate to the degree of localization 
of the states at the band edges - the more they are delocalized the larger is the spatial range within which the 
self-energy correction to TB parameters are to be considered.

\begin{figure}[t]
\includegraphics[scale=0.25]{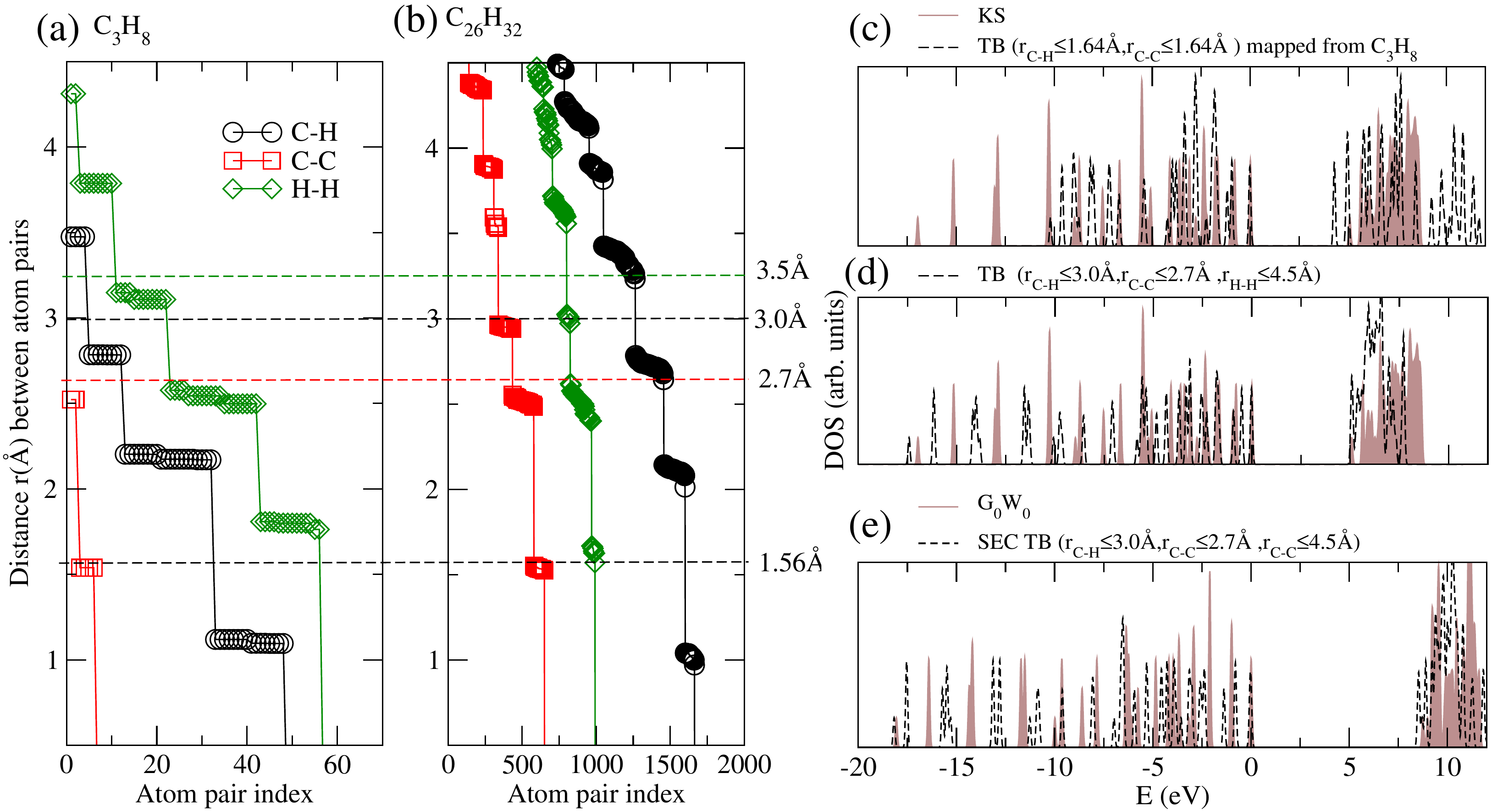}
\caption{Distribution of distance between pairs of atoms in (a): reference (\nds\\) and (b): target (\penta\\) systems. 
(c-d): Match between DFT DOS and mapped TB DOS as  with an  increasing spatial range of neighbourhood considered for mapping. 
(e):  Match between DFT+G$_0$W$_0$ DOS and mapped self-energy corrected TB DOS.}
\label{pentamap}
\end{figure}

\subsubsection{Nano-diamonds}

Fig.\ref{SEChopping} suggests that the extent of SEC to TB parameters are spatially limited mostly within the third nearest 
neighbourhood, implying possible transferability of SE corrected TB parameters to large covalent systems from 
smaller reference systems of which are large enough to accommodate the full spatial range of non-nominal SEC to TB parameters.  
Accordingly, mapping in nano-diamonds is demonstrated with \nds \\ and \ada \\ (adamatane) as reference systems to map to
nano-diamonds \penta \\ (pentamantane) and \ndl\\. 

We start with attempts to map \ada \\, \penta \\ and \ndl \\ targets from \nds \\ reference in \spthree \\ HAO basis.
The mapping process starts with plotting the distance of atom pairs (C-H, C-C, H-H) for target and reference systems.
As seen in Fig.\ref{adamap}(a) there is one-to-one correspondence of C-C bonds between \nds \\ and all targets 
up to approximately 2.5\AA, which is  the second nearest C-C distance.
For C-H and H-H pairs, such correspondence exists up to about 3\AA\ and 3.75\AA \ respectively.
These correspondences decide the range of hopping parameters to be mapped.
Notably, \nds \\ has two varieties of C atoms - one with two(two) C(H) neighbours, 
and the other with one(three) C(H) nearest neighbours, whereas, \ada \\ has C atoms with 
three(one) C(H) neighbours and two(two) C(H) neighbours. 
Additionally, \penta \\ and \ndl \\ have C atoms with all C nearest neighbours(nn). 
Exact match of $\zeta$ between all atoms of reference and target systems is thus impossible in these examples. 
%
Matching  $\zeta$ and $\xi$ will therefore be less effective in mapping from \nds\\.
Also, since there is only one C atom with two(two) C(H) neighbours in \nds\\, matching  $\zeta$ can be restrictive
in terms the variety of orientations.
We thus opt for matching  structural parameters within a tolerance for $\zeta$ set to the minimum difference 
of $\zeta$ values between similar type of atoms in reference and target systems to ensure maximal matching of $\zeta$
besides finer matching of structural parameters. 
%
As obvious, a better choice of reference system than \nds\\ with C atoms having all varieties of neighbourhood 
can be easily made.
However, we deliberately chose to test mapping from C$_3$H$_8$ which is the smallest possible reference system
with just one C atom with two(two) C(H) neighbours, since such C atoms dominates the surfaces of the nano diamonds and
are thereby expected to host the states at the edges of the valence and conduction bands.
Surprisingly, as evident in Fig.\ref{adamap}(c),  with mapping of only the nn-hopping terms from \nds\\ to \ada\\, 
the mapped TB DOS already matches reasonable well with DFT DOS of \ada\\ in terms of the band-gap and DOS around band edges.
With increase in range of hopping to  2.7\AA (nn,2n), 3\AA (nn,2n,3n) and 4\AA (2n,3n,4n)) for C-C, C-H and H-H pairs based on availability
of one-to-one mapping[Fig.\ref{adamap}(a,b)] the match of mapped TB DOS and DFT DOS[Fig.\ref{adamap}(d)] extends deeper into the valence band.
The quality of match improves further with additional mapping of C-H and H-H atom pairs up to 4.5\AA\ [Fig.\ref{adamap}(e)]
without compromising on tolerance factors. 
Notably, the range of hopping of C-H and H-H, although are more than that of C-C, are actually consistent 
with the range of C-C hopping, since the farthest H atoms considered are associated with two second nearest C atoms.   
The same mapping parameters are then used to map self-energy corrected TB (SEC-TB) parameters of \nds\\ to \ada\\ leading to a good match of 
not only the SEC-TB mapped band-gap and the QP band-gap calculated at the G$_0$W$_0$ level, but also the
SE corrected DOS of the valence band[Fig.\ref{adamap}(f)].  

Next we attempt mapping \penta \\ from smaller references, starting with mapping from \nds\\ to \penta\\, 
which is about five times increase in system size.
Mapping of only nearest neighbour C-C and C-H hopping underestimates  band-gap by about 15\%[Fig.\ref{pentamap}(c)].
Mapping all hopping parameters up to  
upto 4.5\AA\, which is the maximum range of hopping available in the reference, drastically improves overall match of not only 
mapped TB DOS and DFT DOS[Fig.\ref{pentamap}(d,e)] but also mapped SEC-TB DOS and DFT+G$_0$W$_0$ DOS[Fig.\ref{pentamap}(f)],
as is seen in case of mapping \ada\\ from \nds\\. 
%

\begin{figure}[t]
\includegraphics[scale=.24]{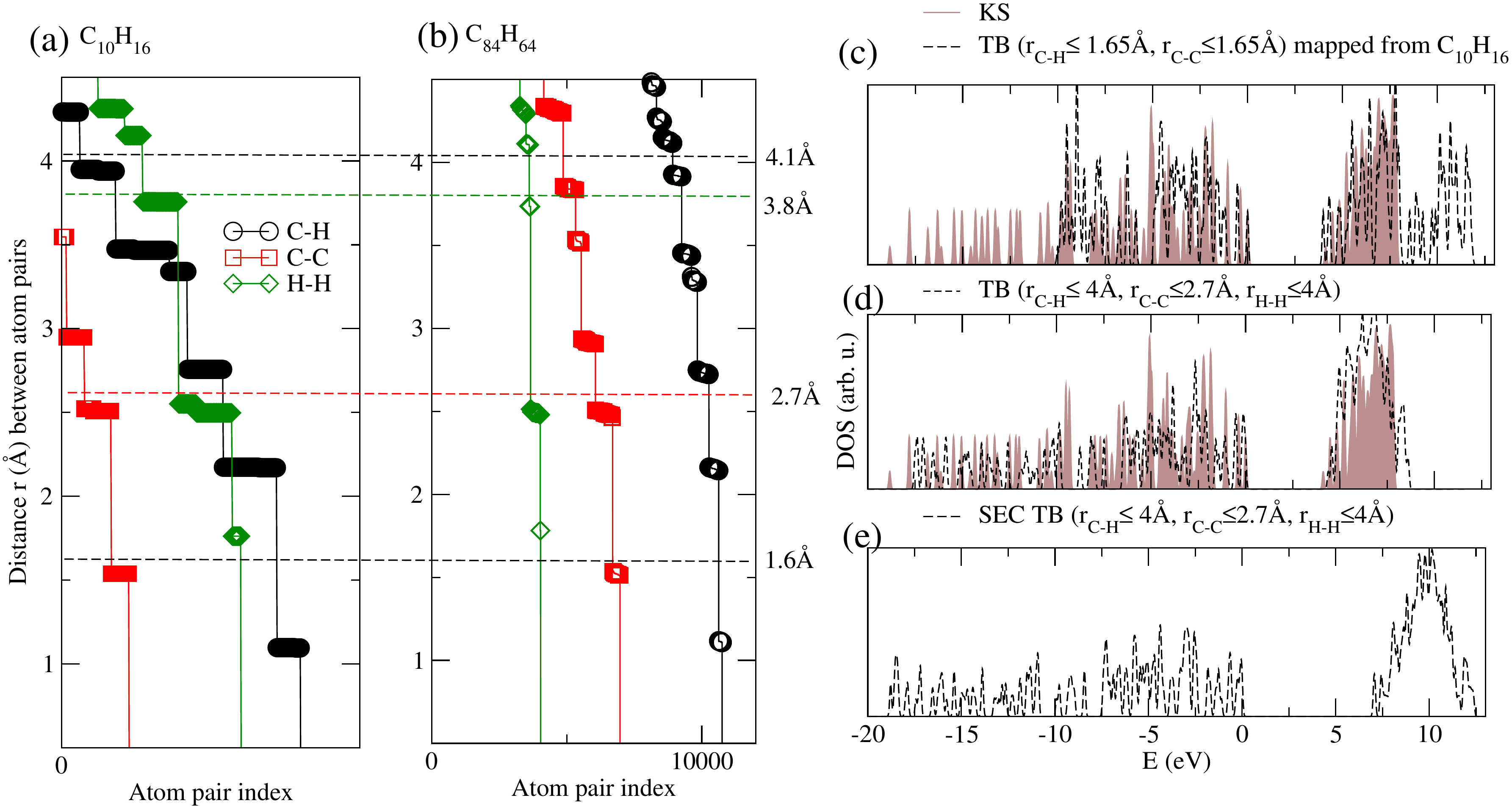}
\caption{Distribution of distance between pairs of atoms in (a): reference (\ada\\) and (b): target (\ndl\\) systems. 
(c-d): Match between DFT DOS and mapped TB DOS as  with an increasing spatial range of neighbourhood considered for mapping. 
(e):  Match between DFT+G$_0$W$_0$ DOS and mapped self-energy corrected TB DOS.}
\label{n148map}
\end{figure}

Finally we demonstrate mapping to \ndl\\ from \ada\\, which is about six time enhancement in system size.
Mapping of only the nearest neighbour C-C and C-H bonds results into good match of
the mapped TB  band-gap[Fig.\ref{n148map}(c)] with the DFT band-gap.
With further mapping of hopping parameters upto 2.75\AA (nn,2n), 4\AA\ (nn,2n+) and 
4\AA\ (2n,3n+)[Fig.\ref{n148map}(a,b)] for C-C, C-H and H-H pairs, satisfactory
match of the entire valence band and a good match[Fig.\ref{n148map}(d)] of the band-gap is achieved.
Mapping of SEC of TB parameters from \ada\\ to \ndl\\ results into a QP band-gap of about 7.2 eV which is within 
5\% deviation from the QP band-gap implied in 
literature\cite{raty2005optical,raty2003quantum,drummond2005electron,sasagawa2008route}.

\begin{figure}[t]
\includegraphics[scale=0.53]{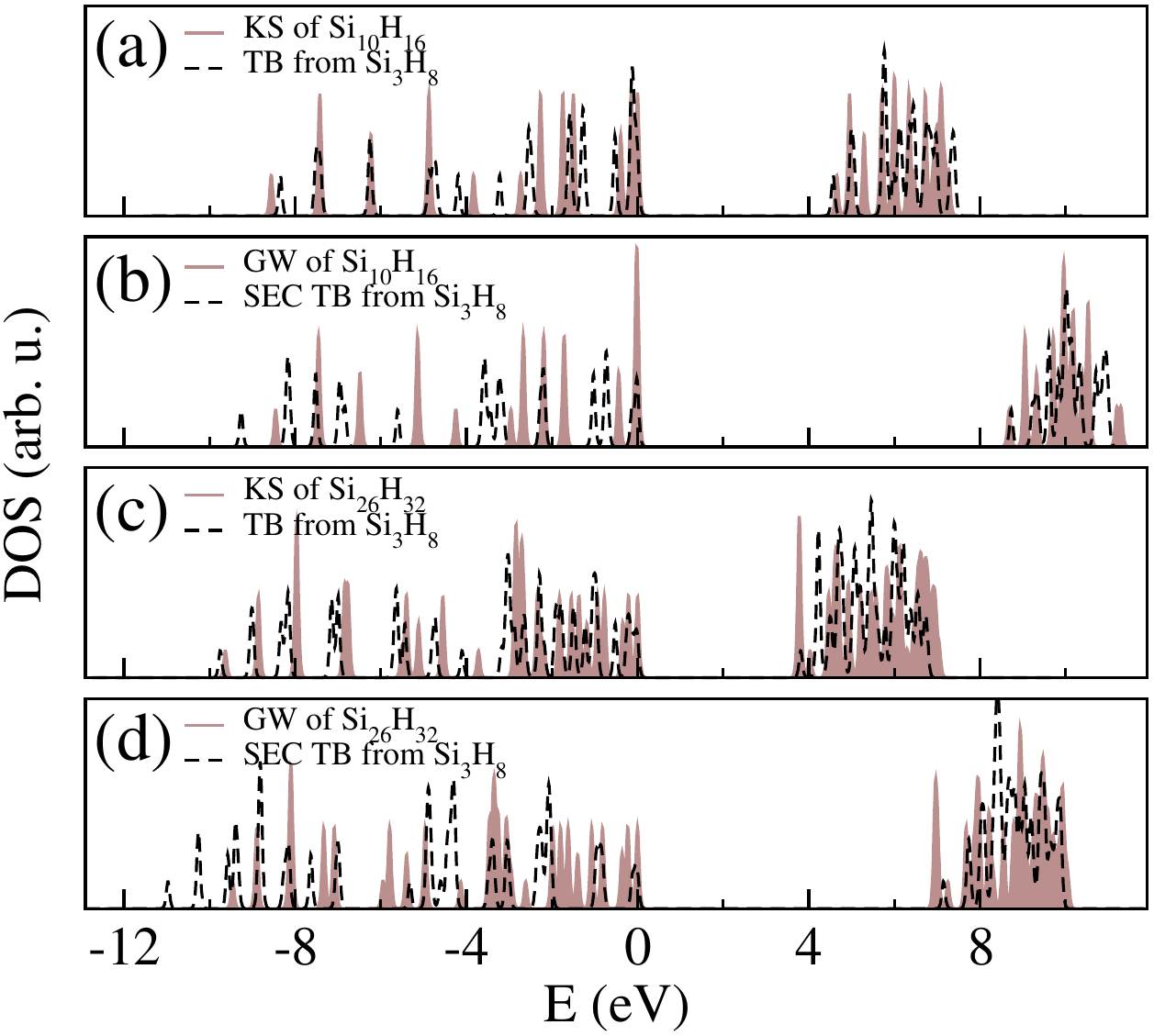}
\caption{
(a,c): Match between DFT DOS and TB DOS with parameters mapped from Si$_3$H$_8$. 
(b,d):  Match between DFT+G$_0$W$_0$ DOS and SEC-TB DOS using mapped self-energy corrected TB parameters from  Si$_3$H$_8$.
}
\label{Sinanodiamond}
\end{figure}

In Fig.\ref{Sinanodiamond} we show similar mapping of TB parameters at the DFT and DFT+G$_0$W$_0$ levels for Si based nano-diamonds.
Like in case  of nano-diamonds, mapping of hopping up to second nearest Si neighbours and H atoms associated with them from Si$_3$H$_8$, 
renders good match of the SEC-TB band-gap with the explicitly estimated DFT+G$_0$W$_0$ band-gap almost up to
six times escalation of system size.
These results imply consistency in transferability of SEC corrected TB parameters with increasing principal quantum number of 
valence orbitals.
\section{Conclusion}
In conclusion, construction of naturalized hybrid atomic orbitals(HAO) is proposed as
the common eigen-states of the non-commuting set of finite first-moment matrices corresponding 
to the orthogonal directions. 
%
%
Hybridization and orientations of HAOs are numerically naturalized as per their anticipated immediate 
atomic neighbourhood. 
%
Choice of gauge based on the HAOs leads to the construction of the 
hybrid atomic Wannier orbitals (HAWO) from Kohn-Sham(KS) single particle states, 
rendering a multi-orbital orthonormal tight-binding(TB) basis
locked to the nearest neighbourhood. 
HAWO basis allows calculation of single TB parameters per bond from first principles,  
and facilitate their easy transfer across iso-structural 
systems through mapping of immediate atomic neighbourhoods and projection of charge centres 
learned in the process of naturalization of the HAOs.
The mapping allow effective bottom-up transfer of self-energy corrected 
TB parameters estimated within the $GW$ approximation of many-body perturbation theory in HAWO basis, 
from smaller reference systems to much larger target systems having similar covalent atomic neighbourhoods,
suggesting a possible route towards  computationally inexpensive estimation of quasi-particle structures of large covalent systems 
within acceptable range of accuracy, 
with extra computational cost scaling as $N^2$, 
beyond the explicit computation of self-energy correction
for smaller reference systems which typically scale as $N^4$.
%
Demonstrated in nano-ribbons and nano-diamond systems, 
the transferability of self-energy corrected multi-orbital TB parameters in HAWO basis,
is rooted at the spatial localization of the extent of self-energy correction predominantly within the third nearest neighbourhood,
which appears to be robust for $\sigma$ bonds but  lesser so with $\pi$ bonds and unpaired electrons.
 
\section{Acknowledgments}
Computations have been performed in computing clusters supported by the Nanomission of the 
Dept. of Sci.\& Tech. and Dept. of Atomic Energy of the Govt. of India.

\bibliographystyle{unsrt}
\bibliography{references}

\end{document}